\documentclass[trackchanges, twocolumn]{aastex7}
\usepackage[makeroom]{cancel}
\usepackage{scrextend} 
\usepackage{bold-extra} 
\usepackage{color}
\usepackage[T1]{fontenc} 
\usepackage{fancyhdr}
\usepackage{ulem}
\usepackage{xcolor}
\usepackage{relsize}
\usepackage{textcomp}
\usepackage{lipsum}
\usepackage{empheq}
\usepackage{float} 
\usepackage{epstopdf, epsfig}
\usepackage{graphicx} 
\usepackage{graphics}
\usepackage{onimage} 
\usepackage{tabu} 
\usepackage{tabulary} 
\usepackage{tabularx} 
\usepackage{booktabs}

\usepackage{framed} 
\usepackage{listings} 
\usepackage{amsmath} 
\usepackage{amssymb} 
\usepackage{mathrsfs}

\usepackage{algorithm} 
\usepackage{algpseudocode} 

\usepackage{tensor}
\usepackage{upgreek}
\usepackage{physics}
\usepackage[toc, xindy, acronym]{glossaries}
\usepackage{natbib}

\usepackage[title]{appendix}
\usepackage[perpage,bottom]{footmisc}
\usepackage{enumitem}
\usepackage{zref-savepos}
\usepackage{cleveref}
\crefformat{figure}{figure~#2#1#3}
\crefformat{appendix}{appendix~#2#1#3}
\crefformat{equation}{(#2#1#3)}
\usepackage{tikz}
\usepackage{circuitikz}
\usetikzlibrary{calc}
\usetikzlibrary{patterns}
\usetikzlibrary{arrows}
\usetikzlibrary{shapes}
\usetikzlibrary{plotmarks}
\usetikzlibrary{decorations.markings}

\usepackage{pgfplots}
\usepgfplotslibrary{polar}
\usepgflibrary{shapes.geometric}
\usepackage[]{pstricks}

\newcommand{\closesymbol}{\!}
\newcommand{\fourier}[1]{\tilde{#1}}  % Fourier-transformed functions
\renewcommand{\fourier}[1]{#1}  % remove the tilde from Fourier-space quantities

\definecolor{googleblue}{HTML}{4285F4}
\definecolor{googlered}{HTML}{DB4437} 
\definecolor{googleyellow}{HTML}{F4B400} 
\definecolor{googlegreen}{HTML}{0F9D58} 
\definecolor{oxfordblue}{RGB}{0, 33, 71}
\definecolor{pinkhighlight}{RGB}{255, 51, 204}
\definecolor{ppplorange}{RGB}{248, 128, 34}
\definecolor{defaultblue}{RGB}{51, 51, 178}

\newcommand{\ignore}[1]{}

\newcommand{\rmd}{\text{d}}
\newcommand{\rmi}{\text{i}}

\ifdefined\rmJ
\renewcommand{\rmJ}{\text{J}}
\else
\newcommand{\rmJ}{\text{J}}
\fi

\ifdefined\rmI
\renewcommand{\rmI}{\text{I}}
\else
\newcommand{\rmI}{\text{I}}
\fi

\newcommand{\rmZed}{\text{Z}}

\newcommand{\rmGamma}{\mathrm{\Gamma}}

\ifdefined\partd
\renewcommand{\partd}[3][]{\frac{{\partial^{#1} #2}}{{\partial #3}^{#1}}}
\else
\newcommand{\partd}[3][]{\frac{{\partial^{#1} #2}}{{\partial #3}^{#1}}}
\fi

\renewcommand{\vec}[1]{\boldsymbol{#1}}
\ifdefined\vec  % Make sure we got bold vectors
\renewcommand{\vec}[1]{\boldsymbol{#1}}
\else
\newcommand{\vec}[1]{\boldsymbol{#1}}
\fi

\ifdefined\div
\renewcommand{\div}{{\vec{\nabla} \cdot}}
\else
\newcommand{\div}{{\vec{\nabla} \cdot}}
\fi

\newcommand{\gpar}{\vec{\nabla}_\parallel}
\newcommand{\gperp}{\vec{\nabla}_\perp}

\ifdefined\vr
\renewcommand{\vr}{{\vec{r}}}
\else
\newcommand{\vr}{{\vec{r}}}
\fi

\newcommand{\vk}{{\vec{k}}}
\newcommand{\vv}{{\vec{v}}}

\newcommand{\kperp}{k_\perp}
\newcommand{\kpar}{k_\parallel}
\newcommand{\vperp}{v_\perp}

\newcommand{\vpar}{v_\parallel}

\newcommand{\intv}[1][3]{{\int \rmd^{#1} \vec{v} \ }}

\newcommand{\s}{s}

\newcommand{\vths}[1][\s]{v_{\text{th}{#1}}}
\newcommand{\vthi}{\vths[i]}
\newcommand{\vthe}{\vths[e]}

\newcommand{\ms}[1][\s]{m_{#1}}

\newcommand{\mi}{\ms[i]}
\newcommand{\rhoi}{\rho_i}
\newcommand{\rhoe}{\rho_e}
\newcommand{\rhos}{\rho_\s}

\newcommand{\mfpi}{\lambda_{\text{mfp}i}}

\newcommand{\taubar}{\bar{\tau}}

\newcommand{\nos}[1][\s]{n_{0#1}}
\newcommand{\noe}{\nos[e]}
\newcommand{\noi}{\nos[i]}
\newcommand{\dns}[1][\s]{\delta n_{#1}}
\newcommand{\dne}{\dns[e]}

\newcommand{\Tos}[1][\s]{T_{0#1}}
\newcommand{\Toe}{\Tos[e]}
\newcommand{\Toi}{\Tos[i]}

\newcommand{\dTpars}[1][\s]{\delta T_{\parallel {#1}}}
\newcommand{\dTpare}{\dTpars[e]}

\newcommand{\foe}{f_{0e}}

\newcommand{\dfe}{\delta \closesymbol f_{e}}

\newcommand{\pbra}[2]{\left\lbrace #1, #2 \right\rbrace}

\newcommand{\phipot}{\phi}  % electrostatic potential
\newcommand{\phipotk}{\fourier{\phipot}_\vk}

\newcommand{\Apar}{{A_\parallel}}

\newcommand{\upars}[1][\s]{u_{\parallel{#1}}}

\newcommand{\upare}{\upars[e]}

\newcommand{\vE}{\vec{E}}
\newcommand{\vB}{\vec{B}}
\renewcommand{\ub}{\vec{b}}

\newcommand{\vdBperp}{{\delta \closesymbol  {\vec{B}}_{\closesymbol\perp}}}

\newcommand{\exb}{\vE\times\vB}

\newcommand{\Omegas}[1][\s]{\Omega_{#1}}

\newcommand{\Omegai}{\Omegas[i]}

\newcommand{\fenergy}{W}
\newcommand{\helicity}{H}
\newcommand{\fluxe}{\varepsilon_\fenergy}
\newcommand{\fluxh}{\varepsilon_\helicity}
\newcommand{\fluxpm}{\varepsilon^\pm}

\newcommand{\fluxp}{{\varepsilon^{+}}}
\newcommand{\fluxm}{{\varepsilon^{-}}}
\newcommand{\forcing}{\sigma_{\varepsilon}}
\newcommand{\imbalance}{\tilde{\sigma}_c}
\newcommand{\uperp}{\vec{u}_\perp}
\newcommand{\thetapm}{{\Theta^\pm}}

\newcommand{\thetap}{{\Theta^+}}
\newcommand{\thetam}{{\Theta^-}}
\newcommand{\thetapmk}{{\Theta_{\vec{k}}^\pm}}
\newcommand{\thetapms}{{\Theta_{\kperp}^\pm}}

\newcommand{\thetapk}{{\Theta_{\vec{k}}^+}}
\newcommand{\thetaps}{{\Theta_{\kperp}^+}}
\newcommand{\thetamk}{{\Theta_{\vec{k}}^-}}
\newcommand{\thetams}{{\Theta_{\kperp}^-}}
\newcommand{\tnl}{t_\text{nl}}
\newcommand{\z}{z}
\newcommand{\zed}{z}
\newcommand{\betacrit}{\beta_e^\text{crit}}

\newcommand{\kperpstar}{k_{\perp}^{*}}
\newcommand{\gm}{g_m}
\newcommand{\vphase}{v_\text{ph}}

\newcommand{\mcrit}{m_\text{cr}}
\newcommand{\nhermite}{M}

\newcommand{\wkin}{\fenergy_{\text{kin.}}}
\newcommand{\zetapm}{\zeta^\pm}
\newcommand{\zetamp}{\zeta^\mp}
\newcommand{\zetapms}{\zeta^\pm_{\kperp}}

\newcommand{\zetaps}{\zeta^+_{\kperp}}
\newcommand{\zetams}{\zeta^-_{\kperp}}
\newcommand{\zetap}{\zeta^+}
\newcommand{\zetam}{\zeta^-}
\newcommand{\gammaeff}{\gamma_{\mathrm{eff}}}
\newcommand{\alphap}{\alpha}
\newcommand{\cons}{c}
\newcommand{\consp}{\cons^+}
\newcommand{\consm}{\cons^-}
\newcommand{\conspm}{\cons^\pm}

\newcommand{\helinj}{\widetilde{\varepsilon}_H}
\newcommand{\cpar}{c_\parallel}

\newcommand{\constonei}{c_{1i}}
\newcommand{\constonee}{c_{1e}}
\newcommand{\constar}{c_*}
\newcommand{\xii}{\xi_i}
\newcommand{\zedpms}{\tilde{\zed}^\pm_{\kperp}}
\newcommand{\zedps}{\tilde{\zed}^+_{\kperp}}

\newcommand{\zedpsouter}{\tilde{\zed}^+_{\kperp^o}}
\newcommand{\zedmsouter}{\tilde{\zed}^-_{\kperp^o}}
\newcommand{\qparae}{Q_{\parallel e}}
\newcommand{\qperpe}{Q_{\perp e}}
\newcommand{\qparai}{Q_{\parallel i}}
\newcommand{\qperpi}{Q_{\perp i}}
\newcommand{\fluxnorm}{\noe\Toe} %\noi m_i c_s^2

\newcommand{\figscale}{0.75}
\newcommand{\figscalecol}{0.95}
\makeatletter
\usepackage{etoolbox}
\patchcmd\H@refstepcounter{\protected@edef}{\protected@xdef}{}{}
\makeatother

\graphicspath{{../figures/}}
\begin{document}

\title{Turbulent heating in collisionless low-beta plasmas: imbalance, Landau damping, and electron-ion energy partition}

\author[orcid=0000-0001-5028-8047, gname=Toby, sname='Adkins']{Toby Adkins}
%\altaffiliation{Kitt Peak National Observatory}
\affiliation{Princeton Plasma Physics Laboratory, Princeton, New Jersey, 08540, USA}
\affiliation{Department of Physics, University of Otago, Dunedin, 9016, New Zealand}
\affiliation{Rudolf Peierls Centre for Theoretical Physics, University of Oxford, Oxford, OX1 3PU, UK }
\email[show]{tadkins@pppl.gov}  

\author[orcid=0000-0002-8327-5848, gname=Romain, sname='Meyrand']{Romain Meyrand} 
\affiliation{Department of Physics \& Astronomy, University of New Hampshire, Durham, NH, USA}
\affiliation{Department of Physics, University of Otago, Dunedin, 9016, New Zealand}
\email{romain.meyrand@otago.ac.nz}

\author[orcid=0000-0001-8479-962X, gname=Jonathan, sname='Squire']{Jonathan Squire} 
\affiliation{Department of Physics, University of Otago, Dunedin, 9016, New Zealand}
\email{jonathan.squire@otago.ac.nz}

%\collaboration{all}{Collaboration}

%% Use the \collaboration command to identify collaborations. This command
%% takes an optional argument that is either a number or the word "all"
%% which tells the compiler how many of the authors above the command to
%% show. For example "\collaboration[all]{(DELVE Collaboration)}" wil include
%% all the authors above this command.
%%
%% Mark off the abstract in the ``abstract'' environment. 
\begin{abstract}

An understanding of how turbulent energy is partitioned between ions and electrons in weakly collisional plasmas is crucial for modelling many astrophysical systems. Using theory and simulations of a four-dimensional reduced model of low-beta gyrokinetics (the `Kinetic Reduced Electron Heating Model'), we investigate the dependence of collisionless heating processes on plasma beta and imbalance (normalised cross-helicity). These parameters are important because they control the helicity barrier, the formation of which divides the parameter space into two distinct regimes with remarkably different properties. In the first, at lower beta and/or imbalance, the absence of a helicity barrier allows the cascade of injected power to proceed to small (perpendicular) scales, but its slow cascade rate makes it susceptible to significant electron Landau damping, in some cases leading to a marked steepening of the magnetic spectra on scales above the ion Larmor radius. In the second, at higher beta and/or imbalance, the helicity barrier halts the cascade, confining electron Landau damping to scales above the steep `transition-range' spectral break, resulting in dominant ion heating. We formulate quantitative models of these processes that compare well to simulations in each regime, and combine them with results of previous studies to construct a simple formula for the electron-ion heating ratio as a function of beta and imbalance. This model predicts a `winner takes all' picture of low-beta plasma heating, where a small change in the fluctuations' properties at large scales (the imbalance) can cause a sudden switch between electron and ion heating.	

\end{abstract}

%% Keywords should appear after the \end{abstract} command. 
%% The AAS Journals now uses Unified Astronomy Thesaurus (UAT) concepts:
%% https://astrothesaurus.org
%% You will be asked to selected these concepts during the submission process
%% but this old "keyword" functionality is maintained in case authors want
%% to include these concepts in their preprints.
%%
%% You can use the \uat command to link your UAT concepts back its source.

%% From the front matter, we move on to the body of the paper.
%% Sections are demarcated by \section and \subsection, respectively.
%% Observe the use of the LaTeX \label
%% command after the \subsection to give a symbolic KEY to the
%% subsection for cross-referencing in a \ref command.
%% You can use LaTeX's \ref and \label commands to keep track of
%% cross-references to sections, equations, tables, and figures.
%% That way, if you change the order of any elements, LaTeX will
%% automatically renumber them.

\section{Introduction}
\label{sec:introduction}
Plasma turbulence is a ubiquitous feature of many space and astrophysical systems, including the solar wind \citep{GS95,bruno13,chen16}, the solar corona \citep{cranmer05,vanballegooijen11}, accretion flows \citep{ichimaru77,quataert99}, and the intracluster medium \citep{takizawa99,kunz22}. Typically, such systems are weakly collisional, in that their characteristic dynamical timescales are often shorter than those associated with inter-particle collisions \citep{boltzmann96,landau65}. As a result, free energy injected into the plasma by large-scale mechanisms often cannot dissipate directly, and thus must be processed by turbulence to the small scales on which heating can occur. Understanding the mechanisms responsible for this heating and the partitioning thereof between the plasma species (ions and electrons) is crucial in constructing accurate predictive models of the (thermo-)dynamics of such systems. For example, this partitioning can influence the radiative properties, macroscopic transport coefficients, and pressure anisotropy of the plasma, and therefore plays a crucial role in determining observable signatures and large-scale behaviour. 

In many such environments -- broadly, any time there exists a spatial asymmetry --- the plasma turbulence is observed to be imbalanced, meaning it is energetically dominated by Alfv\'enic fluctuations propagating in one direction along the equilibrium magnetic field. This imbalance can arise naturally when the plasma is driven by large-scale motions or gradients that preferentially inject fluctuations propagating in one direction, as is the case in, e.g., the solar wind launched from the solar corona \citep{parker65}. The presence, or otherwise, of non-zero imbalance can have significant implications for the heating channels available to a system. In particular, the recent work of \cite{adkins24} demonstrated that imbalanced Alfv\'enic turbulence can manifest two fundamentally different regimes of turbulent heating depending on the size of the plasma beta (the ratio of the thermal to magnetic pressures) relative to some critical value that depends on the energy imbalance in the outer-scale fluctuations. Below this critical value, the injected free energy is allowed to undergo a Kolmogorov-style \citep{K41} cascade to small perpendicular scales, as in standard theories of low-beta Alfv\'enic turbulence \citep{howes08jgr,sch09,howes10,kawazura19,sch19}. Above this critical beta, however, the constraints imposed on the turbulence by the simultaneous conservation of free energy and some generalised helicity gives rise to the so-called `helicity barrier' \citep{meyrand21,squire22,squire23,adkins24,johnston25}, which prevents energy from cascading past the ion Larmor radius $\rhoi$ to reach these small perpendicular scales. This causes the turbulence to grow to large amplitudes, creating fine parallel structure that excites ICW fluctuations and heats the ions, which absorb the majority of the injected power \citep{squire22,squire23,zhang25}, with only a small fraction heating electrons. This helicity-barrier-mediated turbulence has many features that agree with measurements of the low-beta solar wind, including those of the ion velocity distribution function \citep[see, e.g.,][]{marsch06,he15,bowen22,bowen24,mcintyre24}, helicity \citep{huang21,zhao21}, and properties of the steep spectral slopes of the electromagnetic fields in the ‘transition range’ on scales comparable to the ion Larmor radius which have been observed for decades \citep{leamon98,alexandrova09,sahraoui09} and, more recently, by Parker Solar Probe (PSP) \citep{bowen20,duan21,bowen24}. The helicity barrier may also have an important impact on plasmas in other astrophysical contexts, e.g., by altering the emission properties of black-hole accretion flows, with implications for interpreting images from the Event Horizon Telescope \citep{wong24}.

A key limitation of this paradigm arises from the fact that investigations of the helicity barrier in imbalanced solar-wind turbulence have thus far been conducted without accounting for the effects of electron kinetics. This omission is particularly significant given that such plasmas are either weakly collisional or collisionless, meaning that their kinetic six-dimensional phase space allows for the nonlinear transfer of free energy to small scales in velocity space (`phase mixing'), and thus the heating of electrons through Landau damping \citep{landau46}. In this work, we consider electron heating in both balanced and imbalanced Alfv\'enic turbulence. Using equations derived in a low-beta asymptotic limit of gyrokinetics (the `Kinetic Reduced Electron Heating Model', KREHM; \citealt{zocco11,loureiro16viriato}), we demonstrate that, as was the case in \cite{adkins24}, the turbulence, and resultant heating, is divided into two different regimes depending on the value of the electron beta $\beta_e$ relative to some critical value that corresponds to a curve in the beta-imbalance parameter space. Below this critical beta, electron Landau damping reduces the flux of energy arriving to the smallest perpendicular scales, leading to a steepening of the associated perpendicular energy spectra that becomes more pronounced with increasing imbalance due to the decrease in the cascade rate relative to that of Landau damping. We predict, and demonstrate numerically, that at sufficiently high imbalances this steepening can even become comparable to the steep transition-range spectra observed in the presence of the helicity barrier, an effect not captured by standard theories of kinetic, low-beta Alfv\'enic turbulence \citep[see, e.g.,][]{howes08jgr}. Above the critical beta, we show that the helicity barrier is once again active even in the presence of electron kinetics. Assuming critically-balanced fluctuations, we show that the parallel electron heating rate due to Landau damping is dominated by scales above the spectral break, and derive a formula for this heating rate that shows good agreement with simulations. We note that in neither regime do we find a significant effect of the so-called `plasma echo' \citep{gould67,malmberg68} that could otherwise reverse the cascade in velocity space through `phase unmixing' (see, e.g., \citealt{sch16,adkins18}) and thereby suppress parallel electron heating. Finally, we consider the implications of our results for heating in the context of the low-beta solar wind, outlining a model for the electron-ion heating ratio as a function of $\beta_e$ and the normalised cross-helicity $\sigma_c$. This model predicts a `winner takes all' picture of solar-wind heating: depending on the values of the electron beta and the imbalance, the heating is either entirely dominated by the electron channel (below the critical beta) or the ion one (above it) due to the fact that the electron-ion heating ratio is a strong function of imbalance. In particular, we expect dominant ion heating in typical parameter regimes of interest in the solar wind, consistent with a wide array of observations. Our model makes use of the empirically observed scaling for the location of the spectral break in perpendicular wavenumber space $\kperpstar \rhoi \sim (1-\sigma_c)^{1/4}$ \citep{meyrand21,squire23}, which, as a subsidiary result, we explain theoretically.

The remainder of this paper is organised as follows. \Cref{sec:krehm} motivates and outlines the KREHM system of equations used for the theoretical arguments and simulations throughout this work (\cref{sec:model_equations}), before considering their linear dispersion relation (\cref{sec:linear_dispersion_relation}) and nonlinearly conserved invariants (\cref{sec:nonlinear_invariants}). Details of the numerical implementation and simulation parameters are also briefly discussed (\cref{sec:numerical_setup}). \Cref{sec:landau_damping} considers the dynamics of both balanced and imbalanced turbulence within these model equations. We first consider the impact of imbalance on a constant-flux style cascade of energy (\cref{sec:weakened_cascade}), demonstrating significant spectral steepening in the presence of high imbalance. We then investigate electron Landau damping in the presence of the helicity barrier (\cref{sec:barrier_regime}), before then discussing the intermediate regime (\cref{sec:intermediate_regime}). Motivated by our findings, we outline, in \cref{sec:consequences_for_sw_heating}, a model of heating in the context of the low-beta Alfv\'enic solar wind. Readers uninterested in engaging with the underlying theory may skip ahead to \cref{sec:consequences_for_sw_heating}, working backwards where further clarification is required. Our results are then summarised and the implications thereof discussed in \cref{sec:summary}.

\section{Kinetic reduced electron heating model}
\label{sec:krehm}
Fundamental studies of magnetised plasma turbulence are often conducted within the framework of gyrokinetics, which describes low-frequency, low-amplitude fluctuations on perpendicular scales much smaller than those associated with the variation of the local background magnetic field $\vB_0 = B_0 \ub_0$. These fluctuations exhibit a strong spatial anisotropy, with characteristic wavenumbers satisfying $\kpar \ll \kperp$, for components parallel and perpendicular to $\ub_0$, respectively. The gyrokinetic system of equations \citep{howes06, sch09, abel13} can be derived from the Vlasov-Maxwell system by averaging over the fast timescale associated with the frequency $\Omegas$ of the Larmor motion of the particles under the assumption of this spatial anisotropy, which appears to be well-satisfied in the solar wind \citep{chen13,chen16}. In the limit of low plasma-beta, where the ion thermal speed is much smaller than the Alfv\'en speed, further simplifications can be made. The weak coupling between Alfv\'enic and compressive fluctuations in this regime allows ion kinetics to be neglected even at ion-Larmor scales \citep{sch19}. The resulting equations constitute the `Kinetic Reduced Electron Heating Model' (KREHM, \citealt{zocco11, loureiro16viriato}), which has recently been applied to studies of electron heating in turbulence dominated by kinetic Alfv\'en waves \citep{zhou23PNAS, zhou23MNRAS}. KREHM couples the equations of reduced magnetohydrodynamics (RMHD, \citealt{kadomtsev74,strauss76}) and electron reduced magnetohydrodynamics (ERMHD, \citealt{sch09,boldyrev13}) to the kinetic electron physics. While the formal derivation of KREHM assumes that the electron beta $\beta_e = 8\pi \noe \Toe / B_0^2$ is comparable to the electron-ion mass ratio, $\beta_e \sim m_e / m_i$ \citep{zocco11, adkins22}, the model remains valid for $\beta_e \lesssim 1$ more generally, provided the equilibrium temperature ratio of ions to electrons, $\tau \equiv \Toi / \Toe$, is of order unity ($\noe$ is the equilibrium density of electrons). 

\subsection{Model equations}
\label{sec:model_equations}
In KREHM, the electron dynamics are encoded in the drift-kinetic equation
\begin{align}
	\frac{\rmd \dfe}{\rmd t} + \vpar \gpar\dfe  = \frac{e}{\Toe} \left(\frac{1}{c} \frac{\partial \Apar}{\partial t} + \gpar \phipot \right) \vpar \foe  +  C[\dfe],
	\label{eq:drift_kinetic_equation}
\end{align}
which describes the evolution of the perturbed electron distribution function $\dfe = \dfe(\vr, \vv, t)$ around a homogeneous Maxwellian equilibrium $\foe = \foe(\abs{\vv})$. These perturbations are advected by the $\exb$ flow due to the perturbed electrostatic potential~$\phipot$:
\begin{align}
	\frac{\rmd }{\rmd t} = \frac{\partial}{\partial t} + \uperp \cdot \gperp, \quad \uperp = \frac{c}{B_0} \ub_0 \times \gperp \phipot,
	\label{eq:convective_derivative}
\end{align}
while their parallel motion is along the exact magnetic field, viz., including the perturbation of the magnetic-field direction arising from the parallel component of the magnetic-vector potential $\Apar$:
\begin{align}
	\gpar = \ub \cdot \grad =\frac{\partial}{\partial z} + \frac{\vdBperp}{B_0} \cdot \gperp, \quad \vdBperp = - \ub_0 \times \gperp \Apar.
	\label{eq:gpar}
\end{align}
The combination of terms in the brackets on the right-hand side of \cref{eq:drift_kinetic_equation} can be recognised as (minus) the parallel electric field ($-e$ is the electron charge), while the final term is the collision operator. Although our investigations primarily focus on collisionless dynamics, the role of collisions cannot be entirely disregarded: `phase mixing' associated with the parallel-streaming term on the left-hand side of \cref{eq:drift_kinetic_equation} causes the distribution function to develop small scales in velocity space, where the associated velocity gradients can eventually become steep enough to activate collisions no matter how small they are taken to be. We therefore assume that collisions have a negligible effect on $\dfe$ except at these scales, where the collision operator acts as a direct sink of energy.

It is useful both conceptually \citep[see, e.g.,][]{zocco11,adkins22} and numerically \citep[see, e.g.,][]{loureiro16viriato,mandell18,mandell24} to project the velocity-dependence of $\dfe$ onto an appropriate polynomial basis. Given that $\vperp$ does not appear explicitly in \cref{eq:drift_kinetic_equation}, the perpendicular-velocity-dependence of $\dfe$ can be integrated out of the problem, and we expand the remaining parallel-velocity-dependence as:\footnote{We note that this definition of $g_m$ is different to that adopted by \cite{zocco11,loureiro16viriato,zhou23PNAS}, etc., who demand that the first two moments of $\gm$ must vanish, viz., $g_0 = g_1 = 0$. We prefer to include the electron density and parallel-velocity perturbations in our definition of $\gm$, but this choice is purely aesthetic, and does not affect the physics encoded in the overall set of equations.}
\begin{align}
	& g_m(\vr,t)  = \frac{1}{\noe} \intv \frac{H_m(\vpar/\vthe)}{\sqrt{2^m m!}} \dfe(\vr, \vpar, \vperp, t), \label{eq:hermite_transform}\\
	& \dfe(\vr, \vpar, \vperp, t)  = \sum_{m=0}^\infty \frac{H_m(\vpar/\vthe) \foe}{\sqrt{2^m m!}}  g_m(\vr,t),
	\label{eq:hermite_transform_inverse}
\end{align} 
in which $H_m$ are the Hermite polynomials, $\vthe = \sqrt{2 \Toe/m_e}$ is the thermal speed of electrons, with $m_e$ their mass. Applying \cref{eq:hermite_transform} to \cref{eq:drift_kinetic_equation}, and making use of the recurrence and orthogonality relations of the Hermite polynomials \citep[see, e.g.,][]{abramowitz72}, we arrive at the following equation for the Hermite moments of $\dfe$:
\begin{align}
	&\frac{\rmd g_m}{\rmd t} + \frac{\vthe}{\sqrt{2}} \gpar \left(\sqrt{m+1 } g_{m+1} + \sqrt{m} g_{m-1}\right) \nonumber \\
	&=  \frac{e}{\Toe}\left(\frac{1}{c} \frac{\partial \Apar}{\partial t} + \gpar \phipot \right)\frac{\vthe}{\sqrt{2}} \delta_{m,1} + C[g_m].
	\label{eq:gms_original}
\end{align} 
We have thus replaced our kinetic equation \cref{eq:drift_kinetic_equation} with an infinite hierarchy of `fluid' moments \cref{eq:gms_original}. These are coupled to one another by the parallel-streaming term on the left-hand side which, in the linear regime, is responsible for the phase-mixing of perturbations to the high Hermite numbers associated with small scales in velocity space $m \sim (\delta \vpar/\vthe)^{-2}$ \citep{zocco11,kanekar15,parker16}. Perturbations at these scales are then dissipated by the collision operator $C[g_m]$, the route by which energy lost from the lower-order moments due to Landau damping \citep{landau46} becomes thermalised within this formalism.

The electrostatic potential $\phipot$ appearing in \cref{eq:gms_original} is related to the electron-density perturbation $\dne/\noe = g_0$ by quasineutrality:
\begin{align}
	\frac{\dne}{n_{0e}} = \frac{\delta n_i}{n_{0i}} =  - \taubar^{-1} \frac{e \phi}{T_{0e}} \equiv - \frac{Z}{\tau}(1 - \hat{\rmGamma}_0)  \frac{e \phi}{T_{0e}},
	\label{eq:quasineutrality}
\end{align}
where the operator $\hat{\rmGamma}_0$ can be expressed, in Fourier space, in terms of the modified Bessel function of the first kind: $\rmGamma_0 = \rmI_0(\alpha_i) e^{-\alpha_i}$, where $\alpha_i = (\kperp \rhoi)^2/2$. This becomes $1 - \hat{\rmGamma}_0 \approx -\rhoi^2 \gperp^2/2$ at large scales $\kperp \rhoi \ll 1$, and $1 - \hat{\rmGamma}_0 \approx 1$ at small scales $\kperp \rhoi \gg 1$; the former limit is why \cref{eq:quasineutrality} is sometimes referred to as the \textit{gyrokinetic Poisson equation}. Similarly, the parallel magnetic-vector potential $\Apar$ is related to the parallel-velocity perturbation $\upare/\vthe = g_1/\sqrt{2}$ by Amp\'ere's law:
\begin{align}
	-e n_{0e} \upare & = j_\parallel = \frac{c}{4\pi} \vec{b}_0 \cdot \left( \gperp \times \vdBperp\right), \nonumber \\
	 \Rightarrow \quad  \upare &= \frac{c}{4\pi e n_{0e}} \gperp^2 \Apar.
	\label{eq:amperes_law}
\end{align}
The ion thermal speed does not enter into \cref{eq:amperes_law} as it is formally small in the low-beta limit, and so the electrons are uniquely responsible for carrying the plasma current. 

Using \cref{eq:quasineutrality} and \cref{eq:amperes_law}, we can write our system of equations as 
\begin{align}
	&\frac{\rmd}{\rmd t} \taubar^{-1} \frac{e\phi}{T_{0e}} - \frac{c}{4 \pi e n_{0e}} \gpar \gperp^2 \Apar = 0,  \label{eq:phi_equation} \\
	&\frac{\rmd}{\rmd t}\left(\Apar - d_e^2 \gperp^2 \Apar\right) = - c \left[\frac{\partial \phi}{\partial z} + \gpar \left(\taubar^{-1} \phi - \frac{\dTpare}{e}\right)\right], \label{eq:apar_equation}
\end{align}
where the parallel-temperature perturbation $\dTpare/\Toe = \sqrt{2} g_2$ and all higher-order moments are determined from
\begin{align}
	&\frac{\rmd g_m}{\rmd t} + \frac{\vthe}{\sqrt{2}} \gpar \left(\sqrt{m+1 } g_{m+1} + \sqrt{m} g_{m-1}\right) \label{eq:gms_equation} \\
	&\hspace{3cm} =   C[g_m], \quad m \geqslant 2. \nonumber 
\end{align}
We have chosen here to separate out the first two kinetic moments, \cref{eq:phi_equation} and \cref{eq:apar_equation}, which are responsible for the Alfvénic dynamics central to the discussions that follow. 

Together, \cref{eq:phi_equation}, \cref{eq:apar_equation}, and \cref{eq:gms_equation} form a closed system of equations describing the evolution of a strongly-magnetised, low-beta plasma on scales above the electron-Larmor radius $\kperp \rhoe \lesssim 1$. If the electrons are assumed to be isothermal along exact (equilibrium plus perturbed) field lines, then the parallel-temperature perturbation appearing in \cref{eq:apar_equation} vanishes, and the remaining equations are those of isothermal KREHM \citep[see, e.g.,][]{adkins24}. We note, however, that such an approximation is only valid on scales much larger than the electron-inertial scale $\kperp d_e \ll 1$ (see \cref{sec:linear_dispersion_relation}), wherein the equations of isothermal KREHM reduce to those of FLR-MHD \citep{meyrand21}. Nevertheless, given that the isothermal KREHM system simultaneously captures the RMHD, ERMHD, and inertial-Aflv\'en wave \citep{loureiro16,milanese20} regimes, the KREHM system that we consider here is the simplest possible extension to these reduced fluid systems that captures the effects of electron kinetics.

\subsection{Linear dispersion relation}
\label{sec:linear_dispersion_relation}
Linearising and Fourier-transforming \cref{eq:phi_equation}-\cref{eq:gms_equation}, we find the following dispersion relation for the (real) frequency $\omega$ and damping rate $\gamma$:
\begin{align}
	(\omega - \rmi \gamma)^2  - \kpar^2 \vphase^2\left[1 - \frac{\taubar}{1 + \taubar} \frac{\left(\dTpare/\Toe\right)_\vk}{\left(e \phipot/\Toe\right)_\vk}\right] = 0,
	\label{eq:dispersion_relation}
\end{align}
where the Fourier-space amplitude of the parallel-temperature perturbation is given by
\begin{align}
	\left(\frac{\dTpare}{\Toe}\right)_\vk = &-\left(\frac{2\zeta^2}{\kperp^2 d_e^2} - \taubar\right) 	\label{eq:tpare_linear}\\
	& \times \left[1 + 2 \left(\zeta^2 - \frac{1}{2}\right) \left(1 + \zeta \rmZed\right)\right] \taubar^{-1}\left(\frac{e\phipot}{\Toe}\right)_\vk. \nonumber
\end{align}
with $\zeta = (\omega - \rmi \gamma)/\abs{\kpar} \vthe$ the normalised frequency and $\rmZed = \rmZed(\zeta)$ the plasma dispersion function \citep{faddeeva54,fried61}. It is evident from \cref{eq:dispersion_relation} that when there is no coupling to the kinetic hierarchy --- i.e., in the isothermal limit --- the system supports forwards- and backwards-propagating modes of real frequency $\omega = \pm \kpar \vphase(\kperp)$, where the perpendicular-wavenumber-dependent phase velocity is given by the isothermal result \citep{adkins24}:
\begin{align}
	\vphase(\kperp) = \kperp \rhos \left(\frac{1+\taubar}{1 + \kperp^2 d_e^2}\right)^{1/2} v_A.
	\label{eq:phase_velocity}
\end{align}  
In \cref{eq:phase_velocity}, $\rhos = \sqrt{Z/2\tau} \rhoi$ is the ion-sound radius, related to the (thermal) sound speed by $c_s = \rhos/\Omega_i$, and $v_A = B_0/\sqrt{4\pi n_{0i} m_i}$ is the Alfv\'en speed.

\begin{figure}
	
	\centering
	
	\includegraphics[width=\figscalecol\columnwidth]{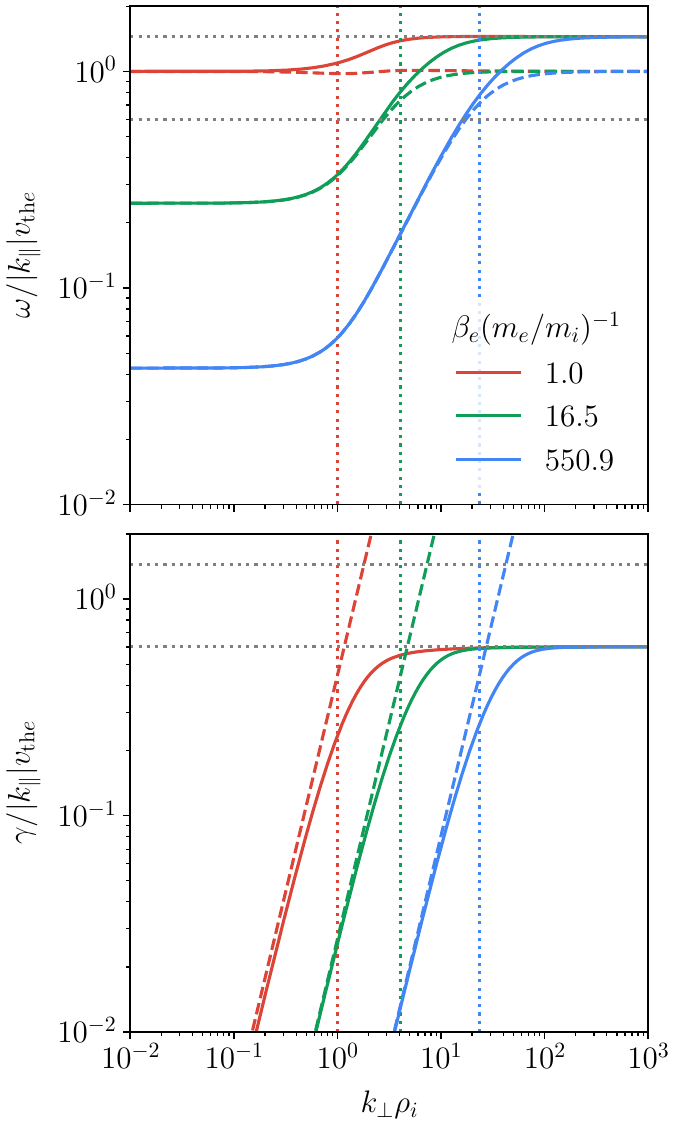}
	
	\caption[]{Solutions to the KREHM dispersion relation \cref{eq:dispersion_relation} normalised to $\abs{\kpar}\vthe$, plotted as a function of perpendicular wavenumber $\kperp \rhoi$, and for $\tau = Z = 1$. The colours indicate different values of $\beta_e (m_e/m_i)^{-1}$. Solid lines are the numerical solutions to \cref{eq:dispersion_relation}, while the dashed lines in the left- and right-hand panels are \cref{eq:phase_velocity} and \cref{eq:gamma}, respectively. The horizontal dotted lines indicate the asymptotic values of the frequency and damping rate as $\kperp \rightarrow \infty$, while the vertical dotted lines indicate $\kperp d_e  = 1$. It is clear that the damping rate is always smaller than the frequency irrespective of perpendicular wavenumber and/or the value of $\beta_e (m_e/m_i)^{-1}$.}
	\label{fig:dispersion_relation}
\end{figure}

On scales much larger than the electron-inertial scale $\kperp d_e \ll 1$, the real part of the numerical solution to \cref{eq:dispersion_relation} is well-approximated by \cref{eq:phase_velocity}, as can be seen from panel (a) of  \cref{fig:dispersion_relation}. In the same limit, the damping rate is given by the kinetic-Alfv\'en-wave (KAW) one \citep[see, e.g.,][]{howes06,zocco11}:
\begin{align}
	\gamma = \frac{\sqrt{\pi}}{4} \abs{\kpar}\vthe \kperp^2 d_e^2 .
	\label{eq:gamma}
\end{align} 
One can readily see from \cref{fig:dispersion_relation}, or indeed from comparing \cref{eq:phase_velocity} and \cref{eq:gamma}, that the damping rate is smaller than the linear frequency at scales $\kperp d_e \lesssim 1$, where $\abs{\gamma}/\omega \sim \kperp d_e /(1+\taubar) \lesssim 1$. It also transpires that this remains true even at scales $\kperp d_e \gtrsim 1$, where both \cref{eq:phase_velocity} and \cref{eq:gamma} cease to be applicable and no analytic solution to \cref{eq:dispersion_relation} can be found. This is, perhaps, an indication that the effects of the kinetic physics could simply be to provide another dissipation channel for the electrons without significantly altering the turbulent dynamics supported by the `fluid' moments \cref{eq:phi_equation} and \cref{eq:apar_equation}. In other words, the dynamics will still be well-described by interactions between waves which, on large scales $\kperp d_e \lesssim 1$, have phase velocities closely approximated by the isothermal result \cref{eq:phase_velocity}. This means that the isothermal limit remains a useful starting point for gathering intuition about the turbulent dynamics \citep{adkins24}. In particular, it will be useful to introduce the eigenfunctions of the forwards- and backwards-propagating modes associated with the isothermal limit. These generalised Els\"asser potentials can be expressed, in Fourier space, as
\begin{align}
	\thetapmk \equiv \sqrt{1 + \kperp^2 d_e^2 }\left[\frac{c}{B_0} \frac{\vphase(\kperp)/v_A}{(\kperp \rhos)^2} \taubar^{-1} \phipotk  \mp \frac{A_{\parallel \vk}}{\sqrt{4 \pi n_{0i}  m_i}}\right],
	\label{eq:thetapm}
\end{align}
and have the property that on the largest scales $\kperp \ll \rhoi^{-1}, d_e^{-1}$, they reduce to the standard RMHD Els\"asser potentials \citep{elsasser50}, viz.,
\begin{align}
	\lim_{\kperp \rightarrow 0} \ub_0 \times \gperp \thetapm = \vec{\zed}^\pm \equiv \uperp \pm \frac{\vdBperp}{\sqrt{4\pi \noi m_i}}. 
\end{align}
We stress that while \cref{eq:thetapm} are not the linear eigenfunctions of the full system of equations \cref{eq:phi_equation}-\cref{eq:gms_equation}, they nevertheless provide a useful basis for our investigation of imbalanced turbulence in KREHM. 

\subsection{Nonlinear invariants}
\label{sec:nonlinear_invariants}
Gyrokinetics conserves the so-called \textit{free energy}, which is the sum of the quadratic norms of the magnetic perturbations, as well as the perturbations of the distributions of both ions and electrons away from equilibrium. Being derived in a subsidiary limit of gyrokinetics, the KREHM system \cref{eq:phi_equation}-\cref{eq:gms_equation} inherits this property, with the free energy taking the form \citep{zocco11,loureiro16viriato,adkins22}:
\begin{align}
	\fenergy = \int \frac{\rmd^3 \vec{r}}{V} &\left[\frac{e^2 n_{0e}}{2 T_{0e}} \left(\phi \taubar^{-1} \phi\right) + \frac{e^2 n_{0e}}{2 T_{0e}} \left(\taubar^{-1} \phi \right)^2 \right. 	\nonumber\\
	& \quad \left. + \frac{ \left| \gperp \Apar\right|^2 + d_e^2 \left(\gperp^2 \Apar \right)^2}{8 \pi} \right] + \wkin, \label{eq:free_energy} 
\end{align}
for a plasma of volume $V$. The first term in \cref{eq:free_energy} is the free energy in the isothermal limit --- consisting of, from left to right, the energies associated with the perturbations of the electrostatic potential, electron density, (perpendicular) magnetic field and parallel electron velocity --- while
\begin{align}
	\wkin = \frac{\noe \Toe}{2} \int \frac{\rmd^3 \vec{r}}{V} \sum_{m=2}^{\infty} \gm^2
	\label{eq:free_energy_kin}
\end{align}
is the energy contained in all of the higher-order kinetic moments. We note that at large scales $\kperp \ll \rhoi^{-1}, d_e^{-1}$, this latter contribution becomes vanishingly small, i.e., $\wkin \ll W$, and we recover from the remainder the usual expression for the free energy in RMHD \citep[see, e.g.,][]{sch09}.

Another quadratic quantity of interest is the \textit{generalised helicity}:
\begin{align}
	\helicity = -\frac{e^2 n_{0e} v_{A}}{cT_{0e}} \int \frac{\rmd^3 \vec{r}}{V} \: \taubar^{-1}\phi  \left(\Apar- d_e^2 \gperp^2 \Apar \right),
	\label{eq:helicity}
\end{align}
which reduces to the MHD cross-helicity at $\kperp \ll \rhoi^{-1}, d_e^{-1}$, and is proportional to the magnetic helicity at $\rhoi^{-1} \ll \kperp \ll d_e^{-1}$. While the free energy \cref{eq:free_energy} is conserved by nonlinear interactions irrespective of the presence, or otherwise, of the kinetic moments, this is not true for the generalised helicity \cref{eq:helicity}. Indeed, we show in \cref{sec:landau_damping} that the helicity is only conserved in the isothermal limit, where the local source/sink due to parallel gradients of the parallel-temperature perturbation vanishes [see \cref{eq:helicity_conservation}]. Nevertheless, we will demonstrate that, under certain conditions, the influence of this helicity source/sink on the overall dynamics becomes negligible, allowing the turbulence to be effectively treated as helicity-conserving.

\subsection{Numerical setup}
\label{sec:numerical_setup}
In what follows, the KREHM system \cref{eq:phi_equation}-\cref{eq:gms_equation} is solved using a modified version of the pseudospectral code TURBO \citep{teaca09} in a triply-periodic box of size $L_x = L_y = L_z = L$ with $n_\perp^2 \times n_z$ Fourier modes. A total of $\nhermite$ Hermite modes are evolved, with $\nhermite=2$ corresponding to the isothermal limit in which only \cref{eq:phi_equation} and \cref{eq:apar_equation} are solved. Although the behaviour of the simulations appears to converge for a relatively low number of Hermite moments (see appendix \ref{app:dependence_on_nhermite}), we opted to be conservative and set $\nhermite = 32$ in the majority of our simulations to ensure adequate resolution in velocity space. Time is measured in units of the parallel Alfv\'en time $t_A = L_z/v_A$. Hyperdissipation is introduced in the perpendicular direction by replacing the time-derivative on the left-hand sides of \cref{eq:phi_equation}-\cref{eq:gms_equation} by
\begin{align}
	\frac{\rmd}{\rmd t} + \nu_\perp \gperp^8.
	\label{eq:hyperdissipation} 
\end{align} 
Hyperdissipation is not employed in the parallel direction as fluctuations at small parallel scales (large parallel wavenumbers) are sufficiently low-amplitude due to the effects of Landau damping. Convergence in velocity space is ensured by introducing hypercollisions of the form
\begin{align}
	C[\gm] = \nu_\text{col.} m^6 \gm, \quad m \geqslant 3,
	\label{eq:hypercollisions}
\end{align}
that model the effects of collisional dissipation.\footnote{Collisions within the KREHM system of equations are often modelled by exploiting the fact that the Hermite polynomials are eigenfunctions of the differential part of the Dougherty \citep{dougherty64} or Lenard-Bernstein \citep{lenard58} collision operators. This gives rise to a collision term of the form $C[\gm] = \nu_\text{col.} m \gm$ for $m \geqslant 3$ \citep{zocco11,loureiro16viriato,mandell18,adkins22,mandell24}. The hypercollisions \cref{eq:hypercollisions} are a straightforward generalisation of this that allows us to capture collisionless dynamics while maximising the range of unaffected scales in velocity space.} The coefficients $\nu_\perp$ and $\nu_\text{col.}$ are adaptive, viz., they are re-evaluated at each timestep to esnure that dissipation occurs near the grid scale, maximising the inertial range in both position and velocity space (details of the numerical implementation can be found in \citealt{meyrand25}). Fluctuations are forced at large scales at $\kperp = 4\pi/L$, $\abs{k_z} = 2\pi/L $ through the form of negative damping  acting on the $\thetapm$ fields \citep{meyrand21}; this method allows the rates of free-energy and helicity injection to be controlled exactly while producing sufficiently random motions to generate turbulence. All of the simulations listed in \cref{tab:simulation_parameters} have $\tau = Z = 1$, though we will retain dependencies on these parameters in analytical expressions for the sake of completeness. 

\begin{table*}
	
%	\centering
	
	\begin{tabular}{l | c  c c c c c c }
		%				\toprule
		&  Resolution & \:\:$\nhermite$ & $\forcing$ & $\rhoi/L$ & $d_e/L$ &$\beta_e (m_e/m_i)^{-1}$ & Sims \vspace*{-0mm}\\
		\hline
		Steepened cascade & $128^3$ & 2 & 0.0 & 0.10 & 0.1000 & 1.0   & 1 \\
		& $128^3$ & 32 & 0.0 & 0.10 & (0.0043, 0.1000) & (1.0, 550.9) & 3 \\
		& $128^3$ & 2  & 0.6 & 0.10 & 0.1000           & 1.0          & 1 \\
		& $128^3$ & 16 & 0.6 & 0.10 & (0.0043, 0.1000) & (1.0, 550.9) & 3 \\
		& $128^3$ & 2  & 0.8 & 0.10 & 0.1000           & 1.0          & 1 \\
		& $128^3$ & 32 & 0.8 & 0.10 & (0.0043, 0.1000) & (1.0, 550.9) & 3 \\
		\hline
		Isothermal restart & $256^3$ & 2  & 0.8 & 0.04 & 0.0017 & 550.9   & 1 \\
		& $256^3$ & 32 & 0.8 & 0.04 & 0.0017 & 550.9 & 9 \\
		\hline
		Beta scan  & $128^3$ & 16 & 0.6 & 0.10 & (0.0043, 0.1000) & (1.0, 550.9)   & 10 \\
		& $128^3$ & 32 & 0.8 & 0.10 & (0.0043, 0.1000) & (1.0, 550.9)   & 10 \\
		
		\hline
		Hermite scan & $128^3$ & (4, 32) & 0.0 & 0.10 & 0.1000 & 1.0   & 8 \\
		& $128^3$ & (4, 32) & 0.0 & 0.10 & 0.0043 & 550.9 & 8 \\
		& $128^3$ & (4, 32) & 0.8 & 0.10 & 0.1000 & 1.0   & 8 \\
		& $128^3$ & (4, 32) & 0.8 & 0.10 & 0.0043 & 550.9 & 8 \\
		\hline
	\end{tabular} 
	
	\vspace{3mm}

	\caption{The parameters used for the KREHM simulations considered in this paper. All simulations have $\tau = Z = 1$. Values in parentheses indicate the minimum and maximum values for the corresponding column, with the final column (`sims') indicating the number of simulations in a given set. Note that a value of $\nhermite=2$ corresponds to an isothermal simulation.}
	\label{tab:simulation_parameters}
	
\end{table*} 

\section{Landau damping in imbalanced Alfv\'enic turbulence}
\label{sec:landau_damping}
Motivated by observations of the solar wind, we consider turbulence that is imbalanced, meaning that it has significantly more energy content in outward-propagating Alfv\'enic structures than in inward-propagating ones. In our system, we associate these outward- and inward-propagating structures with the Elsasser potentials $\thetap$ and $\thetam$, respectively, as, despite not being linear eigenfunctions of the kinetic system, these correspond to $\vec{\zed}^+$ and $\vec{\zed}^-$ fluctuations on the largest perpendicular scales. An imbalanced system naturally possesses some non-zero (generalised) helicity; this is made obvious by writing the (free) energy \cref{eq:free_energy} and (generalised) helicity \cref{eq:helicity} directly in terms of the generalised Els\"asser potentials \cref{eq:thetapm} as
\begin{align}
	\fenergy & = \frac{n_{0i} m_i}{4} \sum_{\vec{k}} \left( \left|\kperp\thetapk\right|^2 + \left|\kperp \thetamk\right|^2 \right) + \wkin, \label{eq:free_energy_thetapm} \\
	\helicity & = \frac{n_{0i} m_i}{4} \sum_{\vec{k}} \frac{ \left|\kperp\thetapk\right|^2 - \left|\kperp \thetamk\right|^2 }{\vphase(\kperp)/v_A}.
	\label{eq:helicity_thetapm}
\end{align}
The associated \textit{energy imbalance} $\imbalance = \helicity/\fenergy$ can be large, with measured values of the normalised cross-helicity (or RMHD imbalance) $\sigma_c = \lim_{{\kperp} \rightarrow 0} \imbalance$ often exceeding $\abs{\sigma_c} \gtrsim 0.9$ \citep{mcmanus20}. This means the effects of non-zero helicity on the turbulence are often significant in the solar wind. 

To investigate such a regime, we consider the case where the energy \cref{eq:free_energy} and helicity \cref{eq:helicity} are injected into our KREHM system at constant rates $\fluxe$ and $\fluxh$ by some large-scale stirring of turbulent fluctuations, denoting the resultant \textit{injection imbalance} --- the ratio of the injected flux of helicity to that of the energy --- as $\sigma_\varepsilon = \abs{\fluxh}/\fluxe$. \cite{adkins24} showed that, under these conditions, the simultaneous conservation of both nonlinear invariants within the isothermal KREHM system gave rise to a critical value of the electron beta
\begin{align}
	\betacrit = \frac{2Z}{1+\tau/Z} \frac{m_e}{m_i} \frac{1}{\forcing^2},
	\label{eq:betacrit}
\end{align} 
separating two dramatically different types of turbulent dynamics. Specifically, in systems where $\beta_e$ was below the threshold given by \cref{eq:betacrit}, the free energy injected at the largest perpendicular scales would undergo an Alfv\'enic cascade down to scales $\kperp \rhoe \lesssim 1$, where it would dissipate, giving rise to the turbulent heating of electrons. Systems with $\beta_e$ exceeding \cref{eq:betacrit}, on the other hand, were unable to support such a constant-flux solution, inevitably forming a helicity barrier \citep{meyrand21,squire22,squire23}. This restricted the cascade of free energy beyond the ion Larmor scale $\kperp \rhoi \sim 1$, leaving most of the injected energy, apart from the balanced component, to accumulate at larger scales $\kperp \rhoi \lesssim 1$, where it would, presumably, eventually contribute to ion heating. Heuristically, the breakdown of the constant-flux solution can be traced to the perpendicular-wavenumber dependence of the phase velocity in the denominator of \cref{eq:helicity_thetapm}. For $\beta_e \gg \betacrit$, the phase velocity increases significantly at scales below the ion Larmor radius (see \cref{fig:dispersion_relation}), creating a disparity in the scaling of the fluxes of the free energy and helicity that cannot be reconciled within a constant-flux framework. At sufficiently low $\beta_e$, however, this increase in phase velocity becomes negligible (see again \cref{fig:dispersion_relation}), allowing a constant-flux solution to be restored.

The findings of \cite{adkins24} did not account for the potential turbulent dissipation channel offered by electron Landau damping, neglected within the isothermal approximation. In particular, as we noted in \cref{sec:nonlinear_invariants}, the inclusion of electron kinetics means that the helicity \cref{eq:helicity} is no longer conserved, with its time derivative now being given by
\begin{align}
	\frac{1}{\noe \Toe}\frac{\rmd \helicity}{\rmd t} = \fluxh + \helinj  + \dots,
	\label{eq:helicity_conservation}
\end{align}
where `$\dots$' stands for terms arising from the perpendicular hyperviscosity \cref{eq:hyperdissipation}, and
\begin{align}
	\helinj = -\frac{e v_A}{\Toe} \int \frac{\rmd^3 \vr}{V} \: \taubar^{-1} \phipot \gpar\frac{\dTpare}{\Toe}
	\label{eq:helicity_injection}
\end{align}
is the rate-of-change of helicity due to the presence of kinetic effects, vanishing in the isothermal limit. We note that this term is sign non-definite, allowing it to act as either a source or sink of helicity. This, combined with the presence of kinetic damping and the associated non-conservation of the `fluid' energy $W-\wkin$, means that it is not obvious \textit{a priori} that the conclusions of \cite{adkins24} will carry over into the kinetic regime, motivating the present study. In the following sections \cref{sec:weakened_cascade} and \cref{sec:barrier_regime}, we conjecture, and numerically verify, theories of electron Landau damping in the $\beta_e \lesssim \betacrit$ and $\beta_e \gg \betacrit$ limits, respectively, before considering the intermediate regime in \cref{sec:intermediate_regime}.

\subsection{Imbalance-steepened cascade}
\label{sec:weakened_cascade}
For values of the electron beta sufficiently below \cref{eq:betacrit}, the isothermal KREHM system is able to support a local, \cite{K41} style cascade that carries the injected flux of energy and helicity from the outer (injection) scale, through some putative inertial range, to the dissipation scale \citep[see section 3.1 of][]{adkins24}. Let us suppose, and verify \textit{a posteriori}, that such a cascade is also supported within the kinetic system. However, Landau damping allows energy to be removed from the cascade at every (perpendicular) scale, meaning that the assumption that the rates of energy injection into the forward- and backward-propagating fluctuations
\begin{align}
	\fluxpm = \frac{\fluxe \pm \fluxh}{2} = \frac{1\pm \sigma_\varepsilon}{2}{\fluxe},
	\label{eq:fluxpm}
\end{align}
will be equal to the associated flux of $\thetapm$ energy throughout the inertial range is no longer valid. Following \cite{howes08jgr,howes11cascade}, we assume that the nonlinear interactions remain local\footnote{The fact that the interactions remain local is not guaranteed; the nonlinearities in \cref{eq:phi_equation}-\cref{eq:gms_equation} readily allow for the coupling between modes at disparate wavenumbers. While some theories of imbalanced turbulence are non-local \citep[see, e.g.,][]{beresnyak08,sch22}, we choose to consider a local theory in order to focus on understanding the effect of the Landau damping on the turbulence, rather than present a comprehensive phenomenology of imbalanced turbulence.} and estimate the spectral energy fluxes $\Pi^\pm(\kperp)$ as
\begin{align}
	\Pi^\pm(\kperp) \sim  \left(\tnl^\pm\right)^{-1} \frac{\left(\kperp \thetapms\right)^2}{c_s^2},
	\label{eq:fluxpm_scale}
\end{align}
where here, and in what follows, $\thetapms$ refers to the characteristic amplitude of the Els\"asser potentials at each scale $\kperp^{-1}$, rather than to the Fourier transform of the field \cref{eq:thetapm}. Formally, $\thetapms$ can be defined by 
\begin{align}
	\frac{\left(\kperp \thetapms \right)^2}{c_s^2} &= \frac{1}{n_{0e}T_{0e}}\int_{\kperp}^\infty \rmd \kperp' \: E_\perp^\pm (\kperp') , 	\label{eq:thetapm_spectra_def} \\
	 E_\perp^\pm(\kperp) &= 2\pi \kperp  \int_{-\infty}^\infty \rmd \kpar \:  \frac{n_{0i}m_i}{4}\left< \left| \kperp \thetapmk \right|^2 \right>, \nonumber
\end{align}
where $E_\perp^\pm(\kperp)$ is the 1D perpendicular energy spectrum of $\thetapm$ [cf. the first term in \cref{eq:free_energy_thetapm}], and in which the brackets denote an ensemble average. An alternative definition would be via a second-order structure function \citep[see, e.g.,][]{davidson13}. Perturbations of other fluctuating quantities will similarly be taken to refer to their characteristic amplitude at a given perpendicular scale. 

\subsubsection{Nonlinear times and critical balance}
\label{sec:nonlinear_times}
In order to proceed, we need an expression for the nonlinear times appearing in \cref{eq:fluxpm_scale}. However, determining these is not a straightforward task, remaining an open research question even in the RMHD regime \citep{sch22}. Indeed, it is the disparity between the amplitudes of the strong $\thetap$ and weak $\thetam$ perturbations observed in highly imbalanced ($\fluxp/\fluxm \gg 1$) turbulence that makes it challenging to construct a self-consistent theory of such turbulence without invoking assumptions that are difficult to rigorously justify. For example, \cite{lithwick07} assumed that both fields undergo local, strongly-turbulent cascades, which had the immediate implication that the ratio of the nonlinear times exhibits the same scaling as the ratio of the field amplitudes, viz.,
\begin{align}
	\frac{\tnl^+}{\tnl^-} \sim \frac{\thetaps}{\thetams} \sim \frac{\Pi^+}{\Pi^-} > 1,
	\label{eq:nonlinear_times_ratio}
\end{align}
where the last estimate follows from combining the first with those for the energy fluxes \cref{eq:fluxpm_scale}. As pointed out by \cite{lithwick07}, this has the counterintuitive implication that the weaker $\thetams$ perturbation, which is advected by the stronger $\thetaps$ perturbation at some faster rate $(\tnl^-)^{-1}$, can nevertheless coherently advect $\thetaps$ at the slower rate $(\tnl^+)^{-1}$. Furthermore, implicit in \cref{eq:nonlinear_times_ratio} is the assumption that the counter-propagating field is the only source of nonlinear advection available. While this is always satisfied in the RMHD regime ($\kperp \rhoi \ll 1$), the dispersive nature of the KREHM linear modes \cref{eq:phase_velocity} on scales below the ion-Larmor radius ($\kperp  \rhoi \gtrsim 1$) makes nonlinear interactions between co-propagating modes possible, allowing, in principle, a turbulent cascade to be supported by a single component $\thetapm$ \citep[][see also appendix \ref{app:kperpstar}]{cho11,kim15,voitenko16}. The extent to which such interactions contribute to the cascade, and whether they enable a steady-state turbulence that is qualitatively different from the RMHD case, remains an open subject of research.

Additional complications arise when considering the parallel coherence of the two fields. In this context, theories often invoke the assumption of critical balance \citep{GS95,GS97,boldyrev05,nazarenko11}: that the characteristic (linear) time associated with propagation along field lines $\sim \omega$ at some parallel scale $\kpar^{-1}$ is comparable to the nonlinear advection rate at each perpendicular one $\kperp^{-1}$. However, it is not immediately clear how such a critical-balance condition should be formulated in the presence of strong imbalance. Given the disparity of nonlinear times implied by \cref{eq:nonlinear_times_ratio}, one possibility would be to take the nonlinear rate to be the faster one, viz.,
\begin{align}
	\omega(\kpar, \kperp) \sim (\tnl^-)^{-1}.
	\label{eq:critical_balance}
\end{align} 
\cite{lithwick07} argue that the parallel wavenumber appearing in \cref{eq:critical_balance} should be the same for both fields because $\thetaps$ perturbations separated by a distance $(\kpar^-)^{-1}$ (the parallel coherence length of the weaker field) are advected by completely spatially decorrelated $\thetams$ perturbations, which would then imprint their parallel coherence length onto the $\thetaps$ ones, implying that $\kpar^- \sim \kpar^+ \sim \kpar$.

Given these complexities (and indeed others; see \S9.1 of \citealt{sch22}), the precise scalings of the fields $\thetapm$ in imbalanced turbulence remains uncertain. This uncertainty is only further exacerbated by the presence of kinetic effects considered here. Rather than attempting to resolve these issues, we will instead proceed by assuming that \cref{eq:nonlinear_times_ratio} and \cref{eq:critical_balance} are a reasonable first approximation to the turbulent dynamics in the absence of any kinetic effects. As such, the theory that we present here makes no illusions of being a comprehensive phenomenological theory of imbalanced turbulence, but should instead be viewed as an exploration of the possible effects of electron Landau damping in such a regime.

\subsubsection{Els\"asser fluxes}
\label{sec:elsasser_fluxes}
With these caveats in mind, let us return to \cref{eq:fluxpm_scale}. Had the energy fluxes therein been constant with perpendicular wavenumber, and equal to their injected values \cref{eq:fluxpm}, we would have been able to obtain an estimate for the fluctuations of the Els\"asser potentials given \cref{eq:nonlinear_times_ratio}. However, since this is not the case, the perpendicular-wavenumber dependence of these fluxes must first be determined by another constraint. We impose this constraint by assuming that they satisfy the evolution equation for the $\thetapm$ perpendicular energy spectra, analogous to the simplest models used for isotropic hydrodynamic turbulence \citep{batchelor53}:
\begin{align}
	\frac{1}{\noe \Toe} \frac{\partial E_\perp^\pm}{\partial t} =  - \frac{\partial \Pi^\pm}{\partial \kperp} - 2 \gamma \frac{E_\perp^\pm}{\noe\Toe} + S^\pm(\kperp).
	\label{eq:spectral_evolution}
\end{align}
Here, $S^\pm (\kperp)$ is a source term satisfying 
\begin{align}
	\int_{0}^\infty \rmd \kperp \: S^\pm(\kperp) = \fluxpm,
	\label{eq:source_condition}
\end{align}
that represents the injection of energy, usually assumed to be localised at some outer scale~$\kperp^o$ that lies on the largest scales within the system.

In writing \cref{eq:spectral_evolution}, we have made a key assumption that the only additional effect of the electron kinetics on the Aflv\'enic cascade is to provide a linear damping of the energy spectrum, with the damping rate $\gamma$ being given by the solution of the linear dispersion relation \cref{eq:dispersion_relation}. This has important, and indeed quite restrictive, consequences for the behaviour of the Hermite moments \cref{eq:gms_equation} that we will now briefly discuss. In particular, it is not a foregone conclusion that the energy injected into the kinetic hierarchy, via the temperature perturbation in \cref{eq:apar_equation}, will undergo phase mixing to small scales in velocity space. This is obvious from the second term on the left-hand side of \cref{eq:gms_equation}: while energy can undergo phase mixing towards higher $m$ through its coupling to the $m+1$ moment, it can also, in principle, undergo phase unmixing towards lower $m$ via its coupling to the $m-1$ moment. While only the former of these is allowed in the linear regime \citep{kanekar15}, nonlinear effects are able to give rise to the \textit{plasma echo} \citep{sch16,adkins18,nastac24} that returns energy from small scales in velocity space via phase unmixing. This can cause kinetic turbulence to resemble fluid turbulence in that it is only allowed to cascade to small spatial scales in order to reach dissipation \citep[see, e.g.,][]{meyrand19}, rather than also having access to Landau damping as another dissipation channel. We will find, however, in agreement with the results of \cite{zhou23PNAS}, that echoes do not seem to play a significant role in the dynamics, insofar as they do not invalidate the assumption of linear damping. This is discussed further \cref{sec:dynamics_in_hermite_space}.

Looking for a steady-state solution of \cref{eq:spectral_evolution}, and using \cref{eq:fluxpm_scale} and \cref{eq:thetapm_spectra_def} to estimate the energy spectra in terms of the energy fluxes, \cref{eq:spectral_evolution} can be written as
\begin{align}
	\frac{\partial \Pi^\pm}{\partial \kperp } + \frac{2\gamma \tnl^\pm}{\kperp} \Pi^\pm = S^\pm .
	\label{eq:fluxpm_equation_initial} 
\end{align}
Then, making use of \cref{eq:nonlinear_times_ratio} and \cref{eq:critical_balance}, \cref{eq:fluxpm_equation_initial} becomes
\begin{align}
	\frac{\partial \Pi^\pm}{\partial \kperp} + \frac{ 2 \conspm\gamma}{\kperp \omega} \frac{\left(\Pi^\pm\right)^2}{\Pi^-} = S^\pm.
	\label{eq:fluxpm_equation}
\end{align}
The added factors $\conspm$ are to account for the fact that both of the estimates \cref{eq:nonlinear_times_ratio} and \cref{eq:critical_balance} ignored factors of order unity. In general, these factors will be functions of the dimensionless parameters of the plasma, but we will here treat them as constant. Assuming that the forcing due to the source $S^\pm$ is localised at $\kperp = \kperp^o$, and noting that the ratio $\gamma/\omega$ is independent of $\kpar$, \cref{eq:fluxpm_equation} can be solved exactly for wavenumbers $\kperp > \kperp^o$:
\begin{align}
	\Pi^-(\kperp) & =  \fluxm \exp\left[- \int_{\kperp^o}^{\kperp} \frac{\rmd \kperp'}{\kperp'} \: 2 \consm \frac{\gamma(\kperp')}{\omega(\kperp')}\right], \label{eq:fluxm_kperp}\\
	\Pi^+(\kperp) & =  \fluxp \left[1 + \int_{\kperp^o}^{\kperp} \frac{\rmd \kperp'}{\kperp'}\: 2 \consp \frac{\gamma(\kperp')}{\omega(\kperp')} \frac{\fluxp}{\Pi^-(\kperp')}\right]^{-1}. \label{eq:fluxp_kperp}
\end{align}
Finally, the perpendicular energy spectra can be estimated in terms of these fluxes as 
\begin{align}
	E_\perp^\pm(\kperp) \sim \left[\frac{\Pi^\pm(\kperp)}{\fluxpm}\right]^{2/3} \left[E_\perp^\pm(\kperp)\right]^\text{iso},
	\label{eq:reduced_spectra}
\end{align}
where $\left[E_\perp^\pm(\kperp)\right]^\text{iso}$ are the perpendicular energy spectra in the isothermal limit, in which the fluxes are constant and equal to their values at the outer scale. In arriving at this expression, we have used \cref{eq:fluxpm_scale} and \cref{eq:thetapm_spectra_def} to relate the perpendicular energy spectra and fluxes to the amplitudes $\thetapms$, relations which, when combined with the assumption \cref{eq:nonlinear_times_ratio}, imply \cref{eq:reduced_spectra}. Note that in the balanced regime where $\fluxp = \fluxm$ and $\consp = \consm$, both \cref{eq:fluxm_kperp} and \cref{eq:fluxp_kperp} are equal to each other and to the balanced solution of \cite{howes08jgr}, given an appropriate choice of the factors $\conspm$. 

Let us take a moment to examine the behaviour of these expressions in detail. Since the damping rate is strictly positive at all perpendicular wavenumbers (see \cref{fig:dispersion_relation}), the flux of $\thetam$ \cref{eq:fluxm_kperp} is a decreasing function of perpendicular wavenumber, which will (exponentially) steepen the resultant spectrum in comparison to the isothermal/constant-flux limit. Interestingly, \cref{eq:fluxm_kperp} depends on the imbalance only through its value at the outer scale $\fluxm$. This is a consequence of the fact that, if \cref{eq:nonlinear_times_ratio} is to be believed, $\thetam$ is always advected by $\thetap$, a field that is either comparable to or larger than it, and whose associated nonlinear rate $(\tnl^-)^{-1}$ is always comparable to the linear frequency $\omega$ [see \cref{eq:critical_balance}], irrespective of imbalance. The same is not true for the $\thetap$ flux \cref{eq:fluxp_kperp}, whose integrand has an explicit dependence on the ratio of the flux of $\thetam$ to the value of the $\thetap$ flux at the outer scale $\fluxp$. This is because as the imbalance increases, the nonlinear rate for the stronger field $(\tnl^+)^{-1}$ becomes increasingly subdominant to the linear frequency $\omega$, rendering the effects of the otherwise weak Landau damping more significant. Any decrease in the flux of $\thetap$ is thus further accentuated at large imbalances $\fluxm \ll \fluxp $, with a similarly accentuated steepening of the spectrum. Why this is inevitable becomes clear if one considers the $\kperp \rightarrow \infty$ behaviour of \cref{eq:fluxm_kperp} and \cref{eq:fluxp_kperp}. In this limit, the ratio $\gamma/\omega$ becomes a constant independent of perpendicular wavenumber (see \cref{fig:dispersion_relation}), from which it is straightforward to show that
\begin{align}
	\Pi^+(\kperp) \sim \Pi^-(\kperp) \propto \fluxm \left(\frac{\kperp}{\kperp^o} \right)^{-1},
	\label{eq:fluxpm_large_k_limit}
\end{align}
viz., both fluxes have the same small-scale asymptotic scaling. Given that, if $\fluxm \ll \fluxp $, the flux of $\thetap$ is much larger at the outer scale than the flux of $\thetam$, this means that the former must decrease \textit{faster} with perpendicular wavenumber than the latter.

\subsubsection{Testing imbalance-steepened cascade theory}
\label{sec:testing_weakened_cascade_theory}
To test these predictions, we consider the set of simulations in \cref{tab:simulation_parameters} labelled `steepened cascade', consisting of three KREHM simulations at different values of $\beta_e (m_e/m_i)^{-1}$, as well as a fourth isothermal KREHM simulation for comparison, in both the balanced ($\sigma_\varepsilon =0.0$) and imbalanced ($\sigma_\varepsilon = 0.6, \: 0.8$) regimes. Unless otherwise stated, all of the data shown from the `steepened cascade' simulations is time-averaged over the last 20\% of the simulation time.

\begin{figure}
	\centering
	
	\includegraphics[width=\figscalecol\columnwidth]{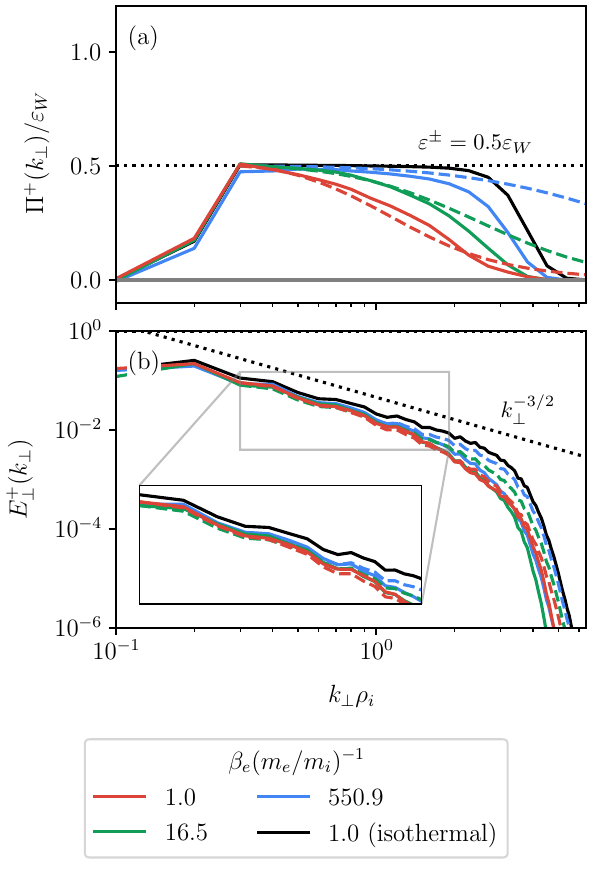}
	
	\caption[]{Time-averaged one-dimensional perpendicular energy fluxes and spectra for the $\thetap$ field from the balanced ($\sigma_\varepsilon = 0.0$) `steepened cascade' simulations in \cref{tab:simulation_parameters}. (a) Energy flux $\Pi^+(\kperp)$ computed directly from the nonlinear terms in \cref{eq:phi_equation} and \cref{eq:apar_equation}, normalised to the total energy flux $\fluxe$. The horizontal dotted line indicates the value of the flux \cref{eq:fluxpm} expected if the system were to maintain a constant-flux cascade. (b) Perpendicular energy spectrum $E_\perp^+(\kperp)$, defined in \cref{eq:thetapm_spectra_def}. The dotted line shows the expected spectral scalings in the isothermal limit \citep[see][]{adkins24}. In both panels, colours indicate the different values of $\beta_e (m_e/m_i)^{-1}$, with solid lines corresponding to the results of direct numerical simulations and dashed lines the theoretical predictions for $\consp = \consm = 2.0$: \cref{eq:fluxm_kperp} or \cref{eq:fluxp_kperp} [panel (a)]; or \cref{eq:reduced_spectra} [panels (b)].}
	\label{fig:weakened_cascade_balanced}
\end{figure}

\begin{figure*}
	\centering
%	\vspace*{-0.5cm}
	\includegraphics[width=\figscale\textwidth]{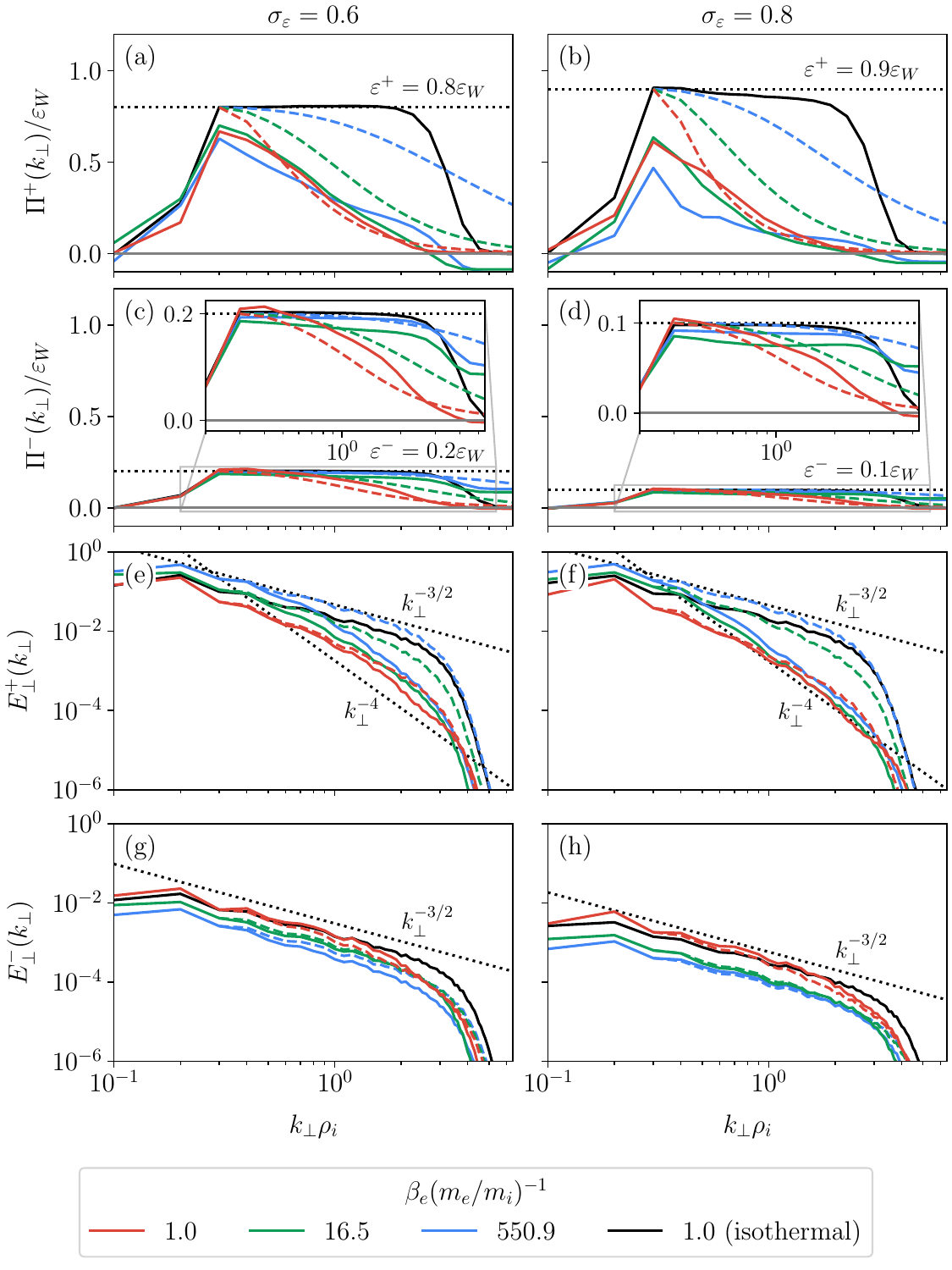}
	
	\caption[]{Time-averaged, one-dimensional perpendicular energy fluxes and spectra for the imbalanced `steepened cascade' simulations in \cref{tab:simulation_parameters}, with the left- and right-hand columns showing $\forcing = 0.6, \: 0.8$, respectively. (a)-(d) Energy fluxes $\Pi^\pm(\kperp)$ computed directly from the nonlinear terms in \cref{eq:phi_equation} and \cref{eq:apar_equation}, normalised to the total energy flux $\fluxe$. The horizontal dotted lines indicate the values of the fluxes \cref{eq:fluxpm} expected if the system were to maintain a constant-flux cascade. (e)-(h) Perpendicular energy spectra $E_\perp^\pm(\kperp)$, defined in \cref{eq:thetapm_spectra_def}. The dotted lines show the expected spectral scalings in the isothermal limit \citep[see][]{adkins24}. In both cases, the colours indicate the different values of $\beta_e (m_e/m_i)^{-1}$, with solid lines corresponding to the results of direct numerical simulations and dashed lines to the theoretical predictions for $\consp = \consm = 2.0$: \cref{eq:fluxp_kperp} [panels (a),(b)]; \cref{eq:fluxm_kperp} [panels (c),(d)]; or \cref{eq:reduced_spectra} (remaining panels). The simulations with $\beta_e (m_e/m_i)^{-1} = 16.5$, and $\beta_e (m_e/m_i)^{-1} = 550.9$ do not conform well to the theoretical predictions as they have formed a helicity barrier (see \cref{sec:intermediate_regime}) and thus do not lie in the relevant regime.}
	\label{fig:weakened_cascade_imbalanced}
\end{figure*}

In figures \labelcref{fig:weakened_cascade_balanced,fig:weakened_cascade_imbalanced}, we plot the time-averaged, one-dimensional perpendicular energy fluxes $\Pi^\pm(\kperp)$ and spectra $E_\perp^\pm(\kperp)$ from these simulations, where the former is calculated directly by summing the contributions of nonlinear transfers above and below a particular $\kperp$ of interest. First, let us focus on the balanced regime, with results for the $\thetap$ field shown in \cref{fig:weakened_cascade_balanced} (the $\thetam$ field exhibits identical behaviour). For the isothermal simulation (black curves), the energy flux of $\thetap$ [panel (a)] is approximately constant (as a function of perpendicular wavenumber), and equal to its injected value \cref{eq:fluxpm}. Note that in this case, the entirety of the injected free energy $\fluxe = \fluxp + \fluxm$ is carried by the turbulence to small scales, where it is then dissipated by the perpendicular hyperdissipation \cref{eq:hyperdissipation}. This is not true for the kinetic simulations, however, in which the measured fluxes appear to be well-reproduced by the theoretical predictions \cref{eq:fluxm_kperp} and \cref{eq:fluxp_kperp}, up to the scale at which the effects of \cref{eq:hyperdissipation} start to become significant. The dependence of this `cascade steepening' on $\beta_e$ is a consequence of the behaviour of the solutions to the linear dispersion relation \cref{eq:dispersion_relation}: while always smaller than the real frequency, the linear damping rate is strongest at scales $\kperp d_e \gtrsim 1$ (see the right-hand panel of \cref{fig:dispersion_relation}), and so we expect its effects to be more pronounced at lower values of $\beta_e$, for which $d_e$ lies on scales comparable to $\rhoi$. This is what we indeed observe, where the simulation with $\beta_e (m_e/m_i)^{-1} = 1.0$ ($d_e/\rhoi = 1.0$, red curves) has fluxes that are significantly more affected by the damping than, e.g., those of the simulation with $\beta_e (m_e/m_i)^{-1} = 550.9$ ($d_e/\rhoi = 0.04$, blue curves). In the latter, the damping rate is significantly smaller than the linear frequency across the full range of resolvable scales. There is also good agreement with the energy spectra \cref{eq:reduced_spectra} up to the dissipation scale [see panel (b) and inset], with departures likely a consequence of the fact that the relationship between the fluxes and spectra implied by \cref{eq:fluxpm_scale} is only approximate.

\Cref{fig:weakened_cascade_imbalanced} shows how the effects of this cascade steepening are enhanced for $\thetap$ in the imbalanced regime for $\forcing =0.6, \: 0.8$. For the simulations with $\beta_e (m_e/m_i)^{-1} = 1.0$ (red curves), the flux of $\thetap$ is reduced almost to zero at scales sufficiently smaller than the outer scale, consistent with the prediction \cref{eq:fluxp_kperp} [panels (a),(b)]. The corresponding spectrum is steepened significantly relative to the isothermal case, reaching a steep $\sim \kperp^{-4}$ scaling at small scales [panels (e),(f)]. The reduction in both the spectrum and flux of $\thetam$ remains comparable to that observed in the balanced case, as expected from \cref{eq:fluxm_kperp}. Conversely, the simulations with higher $\beta_e$ (green and blue curves) diverge markedly from the predictions of \cref{sec:elsasser_fluxes}. The $\thetap$ fluxes are reduced far more than would be expected from the balanced results [cf. panel (a) of \cref{fig:weakened_cascade_balanced}], and the spectra exhibit a pronounced $\sim \kperp^{-4}$ break across $\kperp \rhoi \sim 1$, while the $\thetam$ fluxes are hardly affected. This is to be expected: at the values of the injection imbalance considered here, these simulations have $\beta_e \gtrsim \betacrit$ [$\betacrit(m_e/m_i)^{-1} = 2.78, 1.56$ for $\sigma_\varepsilon = 0.6, 0.8$, see \cref{eq:betacrit}], meaning that they are expected to develop a helicity barrier, placing them in a nonlinear regime entirely different from that which we have been considering in this section (but one that we will discuss in detail in \cref{sec:barrier_regime}).\footnote{Careful readers will have noticed that in the imbalanced regime (\cref{fig:weakened_cascade_imbalanced}), the $\thetap$ fluxes are reduced below the constant-flux value [horizontal dotted lines in panels (a),(b)] even at the outer scale. In the presence of the helicity barrier, this discrepancy arises due to the fact that the system has not yet reached a steady state: although the flux of $\thetap$ arriving to small scales is approximately constant in time, the energy at the largest scales continues to grow. Since the perpendicular energy fluxes measure the nonlinear transfer across a given $\kperp$, we would expect to observe a flux that is less than that which is injected, since a portion thereof must go into increasing the large-scale amplitudes. This is not the case for the simulations with $\beta_e(m_e/m_i)^{-1} = 1.0$, which do not exhibit a helicity barrier. Instead, the reduction in the measured flux arises from the fact that the $d_e$ scale is located sufficiently close to the outer scale that there is damping of the outer-scale fluctuations themselves. It is for this reason that we did not pursue simulations at even lower values of $\beta_e (m_e/m_i)^{-1}$, where this effect would be accentuated. Since this reflects a limitation of the numerical resolution rather than a physical feature of the outer-scale dynamics, we present the theoretical prediction \cref{eq:fluxp_kperp} [dashed lines in \cref{fig:weakened_cascade_imbalanced}(a),(b)] assuming that $\Pi^\pm(\kperp^o) = \fluxpm$.} However, we included these simulations to highlight the following key observation: the spectral steepening induced by electron Landau damping in a Kolmogorov-type cascade can yield spectra that are rather similar to those produced by the helicity barrier. The emergence of steep scalings in both cases means that steep `transition-range' spectra could be observed in the solar wind even in the $\beta_e \ll 1$ regimes where the helicity barrier is not expected to form. We will discuss the implications of this result further in \cref{sec:summary}.

\subsubsection{Dynamics in Hermite space}
\label{sec:dynamics_in_hermite_space}
\begin{figure*}
	\centering
	\includegraphics[width=\figscale\textwidth]{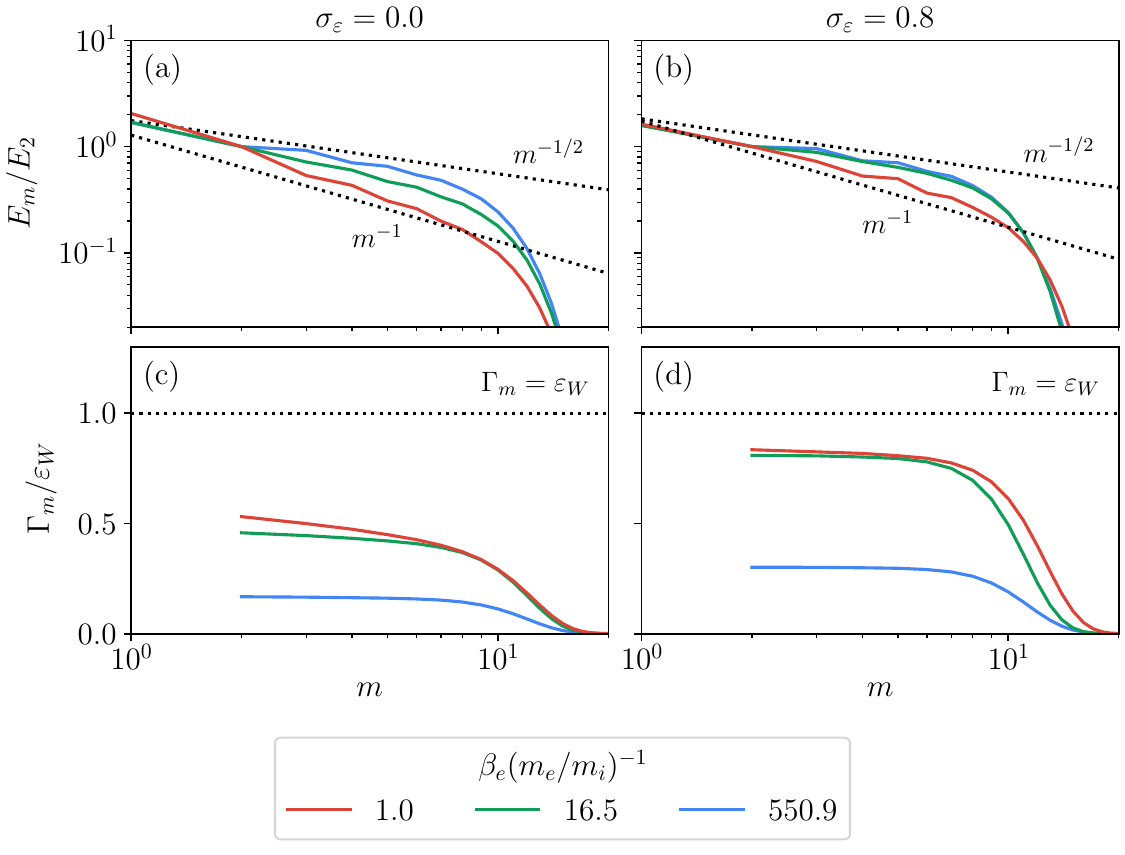}
	
	\caption[]{Time-averaged Hermite spectra and fluxes for the `steepened cascade' simulations in \cref{tab:simulation_parameters}, with the left- and right-hand columns showing the balanced ($\sigma_\varepsilon = 0.0$) and imbalanced ($\sigma_\varepsilon = 0.8$) regimes, respectively. The colours indicate the different values of $\beta_e (m_e/m_i)^{-1}$ (these are same as in figures \labelcref{fig:weakened_cascade_balanced,fig:weakened_cascade_imbalanced}). (a),(b) One-dimensional Hermite spectrum \cref{eq:hermite_spectrum}, normalised to its value at $m=2$, with the dotted lines showing approximate scalings. (c),(d) Hermite flux $\Gamma_m$ normalised to the total energy flux $\fluxe$, the value of latter being indicated by the horizontal dashed line.}
	\label{fig:weakened_cascade_hermite}
\end{figure*}

Before moving onto a more careful analysis of the $\beta_e \gg \betacrit$ regime, we will first examine the dynamics in Hermite space that accompany the behaviour discussed in \cref{sec:testing_weakened_cascade_theory}. It will be useful to introduce the two-dimensional $(\kperp, m)$ spectrum of the energy contained in the Hermite moments:
\begin{align}
	E_{\perp, m}(\kperp) = 2\pi \kperp  \int_{-\infty}^\infty \rmd \kpar \:  \frac{\noe \Toe}{2}\left< \left| g_{m, \vk} \right|^2 \right>,
	\label{eq:km_spectrum}
\end{align}
in which $g_{m,\vk}$ is the Fourier component of the $m$-th Hermite moment, and the brackets once again denote an ensemble average. In the linear regime, it is straightforward to show that the scaling of the one-dimensional Hermite spectrum
\begin{align}
	E_m = \int_{\kperp^o}^\infty \rmd \kperp \: E_{\perp,m}(\kperp),
	\label{eq:hermite_spectrum}
\end{align}
is $E_m \sim m^{-1/2}$ in the large-$m$ limit \citep{zocco11,kanekar15,sch16}. In \cref{fig:weakened_cascade_hermite}(a),(b), we plot \cref{eq:hermite_spectrum} for the kinetic simulations at $\forcing = 0.0, \: 0.8$ shown in figures \labelcref{fig:weakened_cascade_balanced,fig:weakened_cascade_imbalanced}. Those with higher values of the electron beta approximately follow the linear scaling, while the simulations with $\beta_e (m_e/m_i)^{-1} = 1.0$ (red curves) show steeper $\sim m^{-1}$ scalings. This latter deviation from the linear prediction can be understood as follows. The linear Hermite scaling relies on the assumption that the rate of linear phase mixing, approximated in the large-$m$ limit by $\sim \abs{\kpar} \vthe/\sqrt{m} \sim ( \vthe/\sqrt{m})(\kperp^2 \Apar/B_0)$ \citep{zocco11,sch16}, is significantly faster than the nonlinear-advection rate $\sim (c/B_0) (\kperp^2 \phipot)$ [cf. \cref{eq:convective_derivative} and \cref{eq:gpar}, respectively]. As noted by \cite{zhou23PNAS}, this means that at each perpendicular scale there is some Hermite number $\mcrit$ at which the rates of nonlinear advection and phase mixing balance, and above which we expect the effects of nonlinear advection to dominate. Using equipartition between the energies appearing in \cref{eq:free_energy} to relate the amplitudes of $\phipot$ and $\Apar$, it is straightforward to show that this Hermite number is bounded from above by $2\mcrit \lesssim \beta_e (m_e/m_i)^{-1}$ at all perpendicular wavenumbers. We would thus expect the Hermite spectrum to be steepened by the effects of perpendicular advection for the simulations with $\beta_e (m_e/m_i)^{-1} \leqslant 1.0$, which is what is observed. We note, however, that $\mcrit$ was not resolved in the simulations with higher values of $\beta_e(m_e/m_i)^{-1}$ due to computational constraints, meaning that the presence, or otherwise, of echos in this regime remains an open question.

It is important to note that the presence of this steepening does not invalidate our assumption that the kinetic moments act to provide a linear damping of the energy, as described in \cref{eq:spectral_evolution}; the linear scaling of the Hermite spectrum itself is not a necessary condition for this to hold. This becomes evident when considering the one-dimensional flux of energy through Hermite space $\Gamma_m$, which is a direct measure of the rate of energy transfer from the fluid moments \cref{eq:phi_equation} and \cref{eq:apar_equation} into the higher moments in velocity space, viz., the rate of damping of the Els\"asser energy. We compute $\Gamma_m$ directly from the parallel-streaming term in \cref{eq:gms_equation} by summing the contributions of energy transfers above and below a particular $m$ of interest, the results of which are shown in \cref{fig:weakened_cascade_hermite}(c),(d). There is a positive, approximately constant (as a function of $m$) flux of energy from the fluid moments to the high values of $m$ associated with small scales in velocity space, and on which the effects of hypercollisions \cref{eq:hypercollisions} become significant. This is consistent with the picture that all of the energy arriving into the kinetic hierarchy at $m=2$ undergoes nearly unhindered phase mixing to dissipative scales in velocity space; if echoes were playing a significant role in the dynamics, we would expect $\Gamma_m$ to be either significantly steepened, small, or perhaps negative. Furthermore, it is clear from \cref{fig:weakened_cascade_hermite}(c),(d) that a greater fraction of the total energy flux $\fluxe$ is carried to small velocity-space scales at higher imbalance, as would be expected from the predictions of \cref{sec:elsasser_fluxes}; likewise, $\Gamma_m$ is closer to being constant in the imbalanced case, consistent with the slower nonlinear advection of $\thetap$ that renders phase mixing comparatively more efficient (than in the balanced case). However, we must acknowledge a potential caveat: the steepening of $\Gamma_m$ seen for $\beta_e (m_e/m_i)^{-1} = 1.0$ [\cref{fig:weakened_cascade_hermite}(c), red curve] suggests the possibility that in the truly collisionless limit ($\nhermite \rightarrow \infty$), only a small fraction of the energy may ultimately be dissipated at small scales in velocity space. While this remains an open question, computational constraints prevent us from testing it directly. We note that our conclusion that the kinetic moments act as an effective linear damping mechanism may seem to be at odds with the findings of \cite{meyrand19}, who found, in the context of KRMHD \citep{sch09}, that phase mixing was almost entirely suppressed by plasma echo effects. While this difference could be attributed to not resolving $\mcrit$ in most of our simulations, plasma echos were measurably absent in the simulation for which $\mcrit$ \textit{was} resolved [$\beta_e (m_e/m_i)^{-1} = 1.0$], and \cite{zhou23PNAS} reported a similar absence of plasma echos at higher values of $M$, suggesting that plasma echos may in fact be robustly suppressed across the parameter space. Another reason may be the fact that the kinetic component of KRMHD is passive --- in that it is (nonlinearly) advected by the Aflv\'enic dynamics without coupling directly back to it --- which is not the case in KREHM, though a more careful comparative study of these two regimes would be required in order to confirm this.

\subsection{Landau damping in the presence of the helicity barrier}
\label{sec:barrier_regime}
Let us now turn our attention to the $\beta_e \gg \betacrit$ regime, in which we expect the helicity barrier to manifest. Assuming, for the moment, an absence of electron kinetic physics, the helicity barrier exhibits several key characteristics that are observed consistently in both reduced fluid models \citep{meyrand21,adkins24} and hybrid-kinetic simulations \citep{squire22,squire23,zhang25}. Initially, the breakdown of the constant-flux assumption restricts the cascade of free energy to only the balanced portion of the flux (approximately $2 \fluxm$), confining the remainder to large perpendicular scales. The resultant growth in turbulent amplitudes on these scales causes a sharp spectral break to form across the ion Larmor radius, with a steep $\sim \kperp^{-4}$ scaling below it. The perpendicular scale of the break moves towards larger scales in time, with its location in perpendicular-wavenumber space $\kperpstar$ approximated by \citep{meyrand21,squire23}
\begin{align}
	\kperpstar \rhoi \sim \left(1 - \sigma_c\right)^{1/4}. 
	\label{eq:kperpstar}
\end{align}
In appendix \ref{app:kperpstar}, we argue that this scaling arises because the spectral break occurs at the point in perpendicular wavenumber space where co-propagating interactions (i.e., $\thetapm$ with $\thetapm$) begin to be able to compete with the counter-propagating ones. A similar idea was originally proposed by \cite{voitenko16} in an attempt to explain the transition-range spectra often observed in the solar wind, although their resultant theory of spectral scalings relied on the assumption of constant flux, which is impossible to satisfy in the presence of the helicity barrier.

The eventual saturation of the helicity barrier occurs when the energy confined to scales $\kperp \lesssim \kperpstar$ grows to sufficiently large amplitudes that it can access small \textit{parallel} scales and the associated mechanisms of dissipation. In \cite{meyrand21,adkins24}, this saturation was artificial, arising from the use of parallel hyperdissipation, required for numerical convergence. In \cite{squire22,squire23}, saturation occurred via perpendicular heating of the ions by high-frequency ion-cyclotron waves (ICWs) excited at parallel scales $\kpar d_i \sim 1$ that lie outside of the gyrokinetic approximation. However, the inclusion of electron Landau damping in the KREHM system of equations introduces the possibility of saturation \textit{within} the gyrokinetic approximation via electron heating, a possibility not yet studied within the helicity-barrier paradigm.

\subsubsection{Nonlinear heating rate}
\label{sec:effective_damping_rate}
To achieve saturation, there must be a balance between the terms appearing on the right-hand side of the free-energy budget: 
\begin{align}
	\frac{1}{\noe \Toe} \frac{\rmd \fenergy}{\rmd t} = \fluxe  - D_\parallel - D_\perp,
	\label{eq:free_energy_budget}
\end{align}
in which $D_\perp$ is rate of perpendicular (hyper)-dissipation that models the sink of energy at small perpendicular scales, e.g., $\kperp \rhoe \sim 1$, and $D_\parallel$ is the rate of parallel dissipation due to the effects of electron Landau damping. The latter can be written explicitly in terms of the hypercollision operator \cref{eq:hypercollisions} as 
\begin{align}
	D_\parallel = \int \frac{\rmd^3 \vec{r}}{V} \sum_{m=3}^\infty \nu_\text{col.} m^6 \abs{\gm}^2.
	\label{eq:parallel_diss}
\end{align}
As discussed above, the helicity barrier allows only the balanced portion of the energy to be cascaded to small perpendicular scales where it could be dissipated. However, the presence of (even potentially weak) electron Landau damping at all scales means that, in practice, even less energy is dissipated by perpendicular dissipation than in a conventional constant-flux cascade model, viz., it is bounded from above, $D_\perp \lesssim 2 \fluxm \ll\fluxe$. Consequently, if the system is saturated ($\rmd W/\rmd t = 0$), the vast majority of the injected energy flux is expected to be dissipated by $D_\parallel$, for which we will now develop a theory. 

Adopting an analogous approach to \cref{sec:weakened_cascade}, we once again assume that the electron kinetics acts to provide a linear damping of the total energy, which allows us to write the perpendicular spectrum of the parallel dissipation \cref{eq:parallel_diss} as
\begin{align}
	D_\parallel(\kperp) \sim 2\gamma(\kperp) \frac{E_\perp^+(\kperp)}{\noe \Toe},
	\label{eq:parallel_diss_spectrum}
\end{align} 
where $\gamma$ is the linear damping rate, and we have assumed that the turbulence is sufficiently imbalanced that the total energy is dominated by that of the $\thetap$ fluctuations. The parallel-dissipation rate appearing in \cref{eq:free_energy_budget} should, in principle, include contributions from \cref{eq:parallel_diss_spectrum} at all perpendicular scales in the system. However, we assume that contributions from scales below the break $\kperp \gtrsim \kperpstar$ are negligible due to the steep $\sim \kperp^{-4}$ scaling of the $\thetap$ energy spectrum there\footnote{This imposes constraints on the parallel scales of the turbulence. Suppose that at scales $\kperp \gtrsim \kperpstar$, the parallel and perpendicular wavenumbers are related by the scaling $\kpar \sim \kperp^r$, for some positive constant $r$. Then, the damping rate has a scaling with perpendicular wavenumber that is at most as sharp as $\gamma \sim \kperp^{2+r}$, meaning that, from \cref{eq:parallel_diss_spectrum}, the spectrum of the parallel dissipation on these scales will be at least as steep as $D_\parallel \sim \kperp^{-2+r}$. Thus, in order to neglect the contributions from these scales, we require that $r < 1$, a constraint satisfied by the predictions of standard theories of balanced turbulence \citep[see, e.g.,][]{sch09}.}, and thus that the total parallel-dissipation rate is dominated by its contributions from scales \textit{above} the break, viz.,
\begin{align}
	D_\parallel \simeq \int_{\kperp^o}^{\kperpstar} \rmd \kperp \: 2\gamma(\kperp) \frac{E_\perp^+(\kperp)}{\noe \Toe}.
	\label{eq:parallel_diss_integral}
\end{align}

We now estimate $\gamma$, which depends on $\kperp$ both explicitly and through its implicit dependence on $\kpar(\kperp)$. Given that the break lies on scales $\kperpstar \lesssim  \rhoi^{-1} \ll d_e^{-1}$ [see \cref{eq:betacrit}] for $\beta_e \gg \betacrit$, the relevant damping rate appearing in \cref{eq:parallel_diss_integral} will be the KAW one \cref{eq:gamma}. We assume that the parallel wavenumber appearing therein is determined by the critical-balance condition \cref{eq:critical_balance}; though such an assumption is questionable, it appears to be borne out by the agreement seen between our presently developing theory and later numerical simulations. Given that the rate of linear damping is significantly smaller than the linear frequency associated with the phase velocity \cref{eq:phase_velocity} on these scales (see \cref{fig:dispersion_relation}), this linear frequency is taken to be the linear rate appearing in \cref{eq:critical_balance}, viz.,
\begin{align}
	\omega \sim \kpar \vphase \sim \kpar \vthe \kperp d_e (1+\taubar)^{1/2} \sim (\tnl^-)^{-1}.
	\label{eq:linear_frequency}
\end{align}
There is potentially another rate to consider, the phase-mixing rate $\sim \kpar\vthe$ --- this is much faster than the KAW frequency appearing in \cref{eq:linear_frequency}, a fact which perhaps calls into question the validity of this balance. However, a large phase-mixing rate simply implies that any energy that leaves the low-order `fluid' moments \cref{eq:phi_equation} and \cref{eq:apar_equation} associated with the Alfv\'enic dynamics is carried quickly through the higher-order kinetic moments to dissipative scales in velocity space. What matters from the perspective of the Alfv\'enic dynamics is the rate of linear damping, which here is much smaller than the linear frequency in \cref{eq:linear_frequency}.

Next, we estimate the nonlinear time $\tnl^-$ appearing in \eqref{eq:linear_frequency} as the nonlinear $\exb$ advection rate associated with the stronger field. Comparing \cref{eq:convective_derivative} and \cref{eq:thetapm}, neglecting any possible anisotropy in the perpendicular plane, and using the fact that $\taubar/(1+\taubar) \sim 1$ at all perpendicular scales, allows us to rearrange \cref{eq:linear_frequency} for $\kpar \vthe$ as
\begin{align}
	& \kpar \vthe  \sim  \Omega_i\frac{\rhos}{d_e} \left(\kperp \rhos \right)^2 \left(\frac{\thetaps}{\rhos c_s}\right) \nonumber \\
	& \sim  \Omega_i \left(\frac{\rhos}{d_e}\right) \left(\kperp^o \rhos\right)^{(\alphap-1)/2} \left(\kperp \rhos\right)^{(3-\alphap)/2} \left(\frac{W^+}{\noe \Toe}\right)^{1/2}.
	\label{eq:critical_balance_final}
\end{align}
The second estimate in \cref{eq:critical_balance_final} follows from assuming that the perpendicular energy spectrum of $\thetap$ follows $E_\perp^+ \propto \kperp^{-\alphap}$, on scales $\kperp \lesssim \kperpstar$, and that the associated free energy is dominated by its outer-scale contribution [the first estimate in \cref{eq:free_energy_outer}]. Finally, inserting \cref{eq:critical_balance_final} into \cref{eq:gamma}, the damping rate appearing in \cref{eq:parallel_diss_integral} is given by 
\begin{align}
	&\frac{\gamma(\kperp)}{\Omega_i} \nonumber\\
	&= \cpar \left(\frac{d_e}{\rhos}\right) \left(\kperp^o \rhos\right)^{(\alphap-1)/2}\left(\kperp \rhos\right)^{(7-\alphap)/2} \left(\frac{W^+}{\noe\Toe}\right)^{1/2},
	\label{eq:damping_rate_final}
\end{align}
where the added constant $c_\parallel$ is to account for the fact that the scaling estimates \cref{eq:linear_frequency}-\cref{eq:critical_balance_final} ignored factors of order unity. 

Together, \cref{eq:parallel_diss_integral} and \cref{eq:damping_rate_final} constitute our theoretical prediction for the rate at which the free energy \cref{eq:free_energy} is dissipated by electron Landau damping in the presence of the helicity barrier. For typical values of $\alphap$, \cref{eq:damping_rate_final} is an increasing function of $\kperp$, meaning that the integral in \cref{eq:parallel_diss_integral} will be dominated by its value at the upper bound $\kperpstar$. However, given that approximating the integral by this value, as we later do in \cref{sec:electron_heating_rate}, requires there to be sufficient separation between the outer scale $\kperp^o$ and $\kperpstar$ --- a range that gets increasingly small as the imbalance increases [see \cref{eq:kperpstar}] and is necessarily limited by numerical resolution --- we retain the integral in \cref{eq:parallel_diss_integral} for the purposes of our comparisons to simulations. 

\subsubsection{Dependence on amplitude}
\label{sec:dependence_on_amplitude}
\begin{figure}
	\centering
	
	\begin{tikzonimage}[width=\figscalecol\columnwidth]{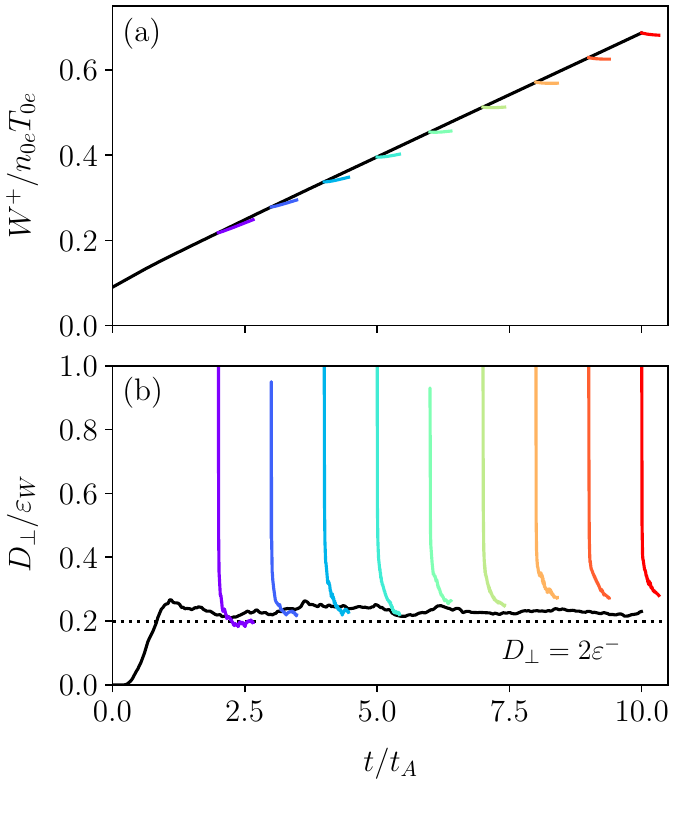}
	\end{tikzonimage}
	
	\caption[]{Timetraces of: (a) the energy $W^+/\noe \Toe$ of the $\thetap$ fluctuations; and (b) the perpendicular dissipation rate normalised to the total energy flux $\fluxe$ for the `isothermal restart' simulations in \cref{tab:simulation_parameters}. The black lines indicate the isothermal simulation, while the colours correspond to each of the individual kinetic simulations. The horizontal dotted line in panel (b) indicates the value at which the rate of perpendicular dissipation is equal to the balanced portion of the injected energy flux, viz., $D_\perp = 2\fluxm$.}  
	
	\label{fig:restart_timetraces}
\end{figure}

\begin{figure*}
	\centering
	
	\begin{tikzonimage}[width=\figscale\textwidth]{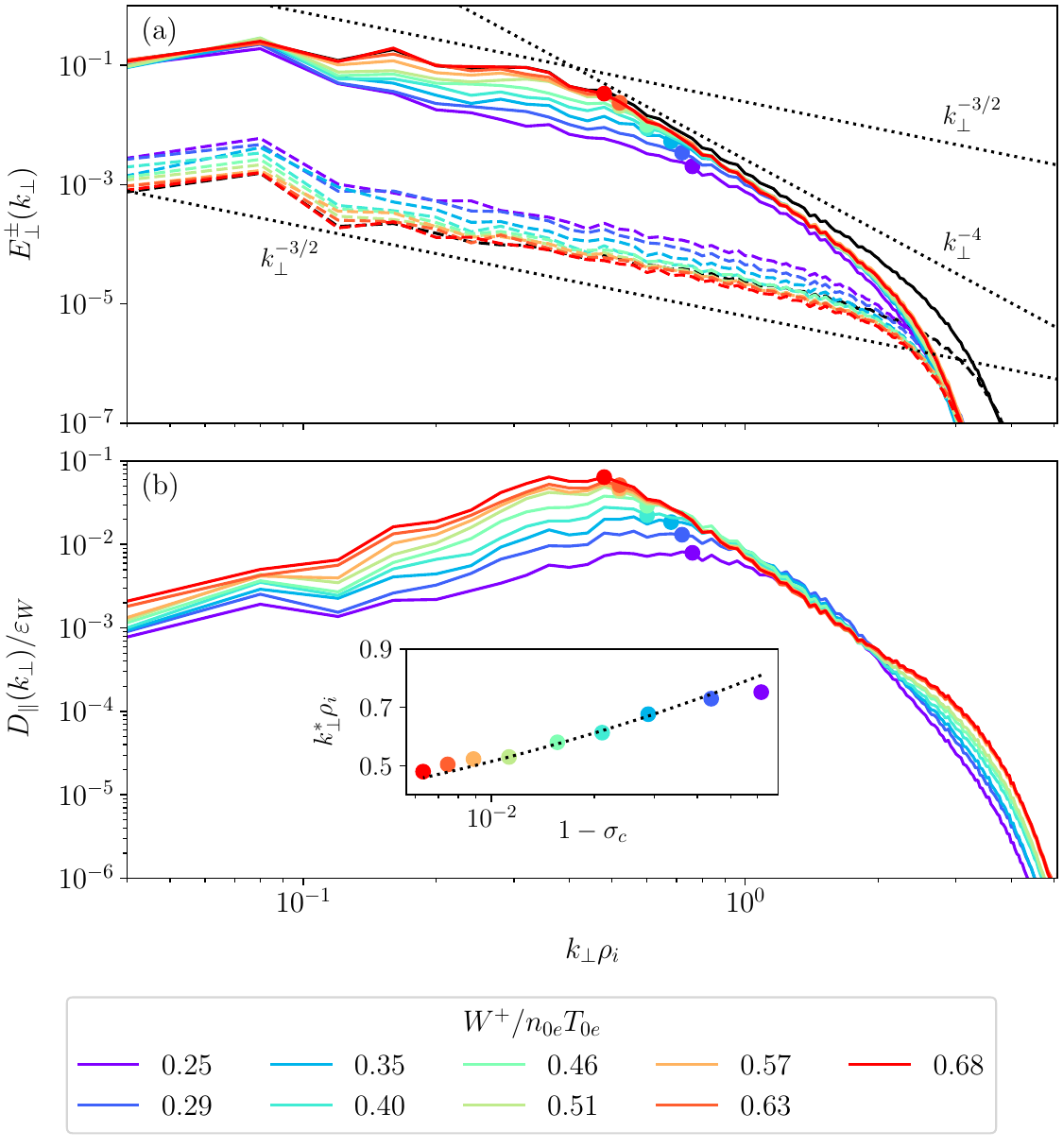}
		%		\node[scale=1.0] at (0.51, 0.38) {\cref{eq:kperpstar}};
	\end{tikzonimage}
	
	\caption[]{Spectra for the 'isothermal restart' simulations in \cref{tab:simulation_parameters}, with the colours the same as those in \cref{fig:restart_timetraces}. (a) One-dimensional perpendicular energy spectra \cref{eq:thetapm_spectra_def}, with the solid and dashed lines corresponding to $E_\perp^+(\kperp)$ and $E_\perp^-(\kperp)$, respectively, and the dotted lines showing expected scalings in the isothermal limit \citep[see][]{adkins24}. The black lines show the spectra of the isothermal simulation at $t/t_A = 10$. The coloured points indicate the value of the $\thetap$ spectrum at the measured position of the spectral break $\kperp =\kperpstar$. (b) One-dimensional perpendicular spectra of the parallel-dissipation rate $D_\parallel(\kperp)$, normalised to the total energy flux $\fluxe$. The coloured points indicate the value of the dissipation spectrum at each of the measured break positions from panel (a). The inset panel plots these as a function of $1 - \sigma_c$, with the dotted line the scaling \cref{eq:kperpstar}.}
	\label{fig:restart_spectra_perp}
\end{figure*}

\begin{figure}
	\centering
	\includegraphics[width=\figscalecol\columnwidth]{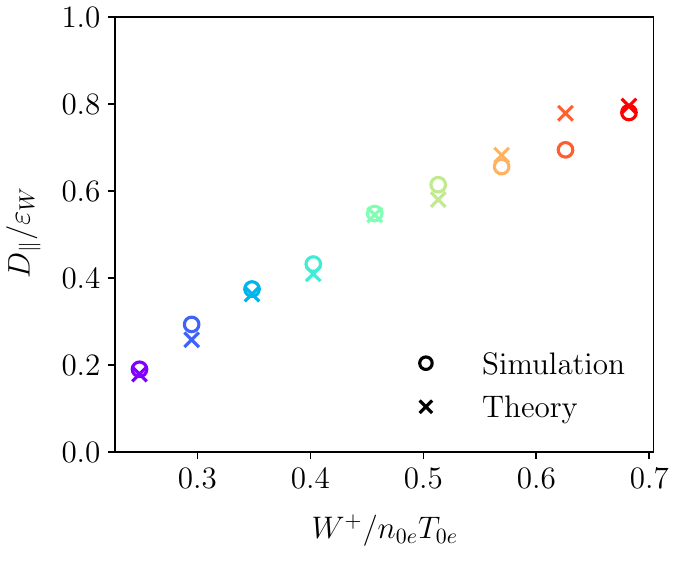}
	
	\caption[]{The parallel-dissipation rate $D_\parallel$ \cref{eq:parallel_diss} for the kinetic `isothermal restart' simulations in \cref{tab:simulation_parameters}, normalised to the total energy flux $\fluxe$, as a function of the energy in the $\thetap$ fluctuations. The colours are the same as those in \cref{fig:restart_timetraces}. The circles and crosses correspond to the measured and theoretical values, respectively, with the latter given by \cref{eq:parallel_diss_integral} for $\cpar = 4.2$ in \cref{eq:damping_rate_final}.}
	\label{fig:restart_heating_rates}
\end{figure}

The parallel-dissipation rate \cref{eq:parallel_diss_integral} depends on the turbulent amplitudes through the explicit presence of the $\thetap$ spectrum in the integrand, as well as through the implicit dependence of $\kperpstar$ on the imbalance. In order to test these predictions over a range of amplitudes, we consider a set of simulations, labelled `isothermal restart' in \cref{tab:simulation_parameters}, consisting of nine kinetic KREHM simulations that were restarted from a single isothermal one. The isothermal simulation was run up to $t/t_A = 10$, with the energy growing in time due to the presence of the helicity barrier  [\cref{fig:restart_timetraces}(a), black line], and the perpendicular dissipation rate roughly equal to the balanced portion of the injected energy flux, that which is allowed through the barrier ($D_\perp \approx 2\fluxm$). The growth of $W^+$ is nearly linear (in time) because there is no parallel dissipation in the isothermal simulations, as discussed in \cref{sec:numerical_setup}. Each of the kinetic simulations are restarted with $g_m = 0$ for $m \geqslant 2$ at unit intervals from $t/t_A = 2$ to $t/t_A = 10$, and run until the perpendicular dissipation rate reached a level comparable to that before the restart, as in \cref{fig:restart_timetraces}(b). The large spikes in the perpendicular dissipation rate seen here are initial transients due to the fact that restart adds $(\nhermite-1)\times n_\perp^2 \times n_z$ new Fourier modes into the simulation. All of the following data is shown from the final time in each restart. This restart method was used due to the prohibitive computational cost of running high-resolution kinetic simulations to the long times required in order to see the dependence of \cref{eq:damping_rate_final} on turbulent amplitude. All of the simulations have $\beta_e (m_e/m_i)^{-1} = 550.9$ ($\beta_e = 0.3$).

In \cref{fig:restart_spectra_perp}(a), we plot the one-dimensional perpendicular energy spectra $E^\pm_\perp(\kperp)$ for each of the kinetic simulations. There is a clear break in the $\thetap$ spectrum between the $\sim \kperp^{-3/2}$ and $\sim \kperp^{-4}$ regimes. The perpendicular wavenumber of the break $\kperpstar$, indicated by the coloured points in \cref{fig:restart_spectra_perp}, moves towards larger scales with time, and its scaling with imbalance appears to conform well to the one expected \cref{eq:kperpstar}, as can be seen from the inset in panel (b). We have taken the break position to be the perpendicular wavenumber at which the measured spectral scaling exponent passes through the midpoint between the $-3/2$ and $-4$ isothermal ones.\footnote{The spectral slopes are determined by performing local logarithmic polynomial fits to the power spectrum over a sliding window in wavenumber space. The resulting slope values are then smoothed using a Gaussian filter to reduce noise and ensure a more robust estimate of the spectral exponent.} Panel (b) shows the one-dimensional perpendicular spectrum of the parallel-dissipation rate $D_\parallel(\kperp)$, normalised to the total energy input $\fluxe$. The coloured points are taken from the aforementioned spectral fit from panel (a). It is clear that the peak of this dissipation occurs around the same scale as the break, i.e., at $\kperp = \kperpstar$, with the dissipation spectrum exhibiting a steep scaling at smaller scales. This is consistent with our expectation that contributions to the overall parallel-dissipation rate from scales $\kperp \gtrsim \kperpstar$ are smaller than those on larger scales. We note that while the $\thetap$ spectrum is slightly steeper than in the isothermal case in the $\kperp > \kperpstar$ transition range because Landau damping is not entirely negligible there, the impact our theory is minimal given these scales contain an insignificant fraction of the total energy. Indeed, \cref{fig:restart_heating_rates} shows good agreement between the measured and theoretical parallel-dissipation rates, the latter being given by \cref{eq:parallel_diss_integral}, in which we have set $\alpha = 3/2$ and $\cpar = 4.2$ in \cref{eq:damping_rate_final}, and evaluated the integral directly from the measured spectrum. Although obtaining exact agreement depends on this value of $\cpar$, it is clear that the theory reproduces the measured scaling of the parallel-dissipation rate with the free energy of the $\thetap$ fluctuations, supporting the assertion that the effect of the electron Landau damping is to damp the turbulence at the linear rate.

\begin{figure}
	\centering
	\includegraphics[width=\figscalecol\columnwidth]{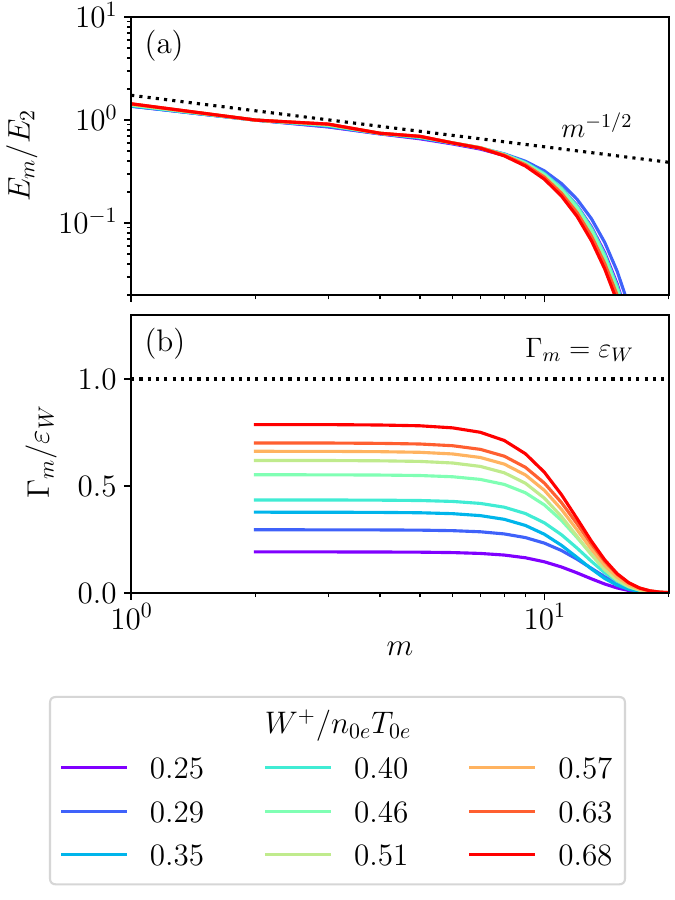}
	
	\caption[]{Hermite spectra and fluxes for the `isothermal restart' simulations in \cref{tab:simulation_parameters}, with the colours once again the same as in \cref{fig:restart_timetraces}. (a) One-dimensional Hermite spectrum \cref{eq:hermite_spectrum}, normalised to its value at $m=2$, with the dotted line showing the linear scaling. (b) Hermite flux $\Gamma_m$ normalised to the total energy flux $\fluxe$, the value of the latter being indicated by a horizontal dashed line.}
	\label{fig:restart_hermite}
\end{figure}

The dynamics in Hermite space also appear to support these assumptions. In \cref{fig:restart_hermite}(a), we plot the one-dimensional Hermite spectrum \cref{eq:hermite_spectrum}, which follows the linear $\sim m^{-1/2}$ scaling for all of the simulations. Unlike in \cref{sec:dynamics_in_hermite_space}, there is no evidence of any (nonlinear) steepening on the Hermite spectrum, which is consistent with being in the $\beta_e \gg \betacrit$ limit [see the discussion following \cref{eq:hermite_spectrum}]. Additionally, \cref{fig:restart_hermite}(b) shows a positive, approximately constant (as a function of $m$) flux of energy from the fluid moments to the high values of $m$ where it dissipates on the hypercollisions \cref{eq:hypercollisions}: all of the energy injected into the kinetic hierarchy at $m=2$ undergoes unhindered phase mixing to dissipative scales in velocity space. As expected, the simulations restarted at higher amplitudes have a higher Hermite flux $\Gamma_m$, and thus a greater overall parallel-dissipation rate (see \cref{fig:restart_heating_rates})

\subsubsection{Dependence on beta}
\label{sec:dependence_on_electron_beta}
The only explicit dependence of the predicted parallel-dissipation rate \cref{eq:parallel_diss_integral} on $\beta_e$ comes from the damping rate \cref{eq:damping_rate_final}, suggesting that the former should inherit the same scaling, viz., $D_\parallel \propto 1/\sqrt{\beta_e}$ [recall that $d_e/\rhos = \sqrt{2Zm_e/(\beta_e m_i)}$]. However, such a scaling cannot be tested directly in our simulations. As discussed in \cref{sec:effective_damping_rate}, the presence of the helicity barrier constrains the fraction of the total injected energy flux $\fluxe$ that can be dissipated at small perpendicular scales, limiting it to the balanced portion, viz., $D_\perp \lesssim 2 \fluxm$.  Consequently, at saturation the remaining energy flux must be dissipated through electron Landau damping, meaning that we will always find (numerically) that $D_\parallel \gtrsim \fluxe - 2\fluxm$, independent of the value of $\beta_e$.  This constraint effectively introduces an implicit $\beta_e$ dependence into the value of $\kperpstar$ in the upper limit of the integral in \cref{eq:parallel_diss_integral} --- to maintain $D_\parallel \gtrsim \fluxe - 2\fluxm$, the large-scale $\thetap$ amplitudes must grow sufficiently, increasing the imbalance and causing $\kperpstar$ to shift towards larger scales, irrespective of the damping rate \cref{eq:damping_rate_final}. Note that this would not be the case in a real physical system due to the existence of other dissipation channels, e.g., perpendicular ion heating due to ICWs, which we discuss further in \cref{sec:consequences_for_sw_heating}. Given that we cannot directly measure the beta dependence of the parallel-dissipation rate, we propose to instead consider an effective damping rate [cf. \cref{eq:parallel_diss_spectrum} and \cref{eq:parallel_diss_integral}]:
\begin{align}
	\gammaeff = \left[\int_{\kperp^o}^{\kperpstar} \rmd\kperp \: D_\parallel(\kperp)\right] \Bigg/ \left[2 \int_{\kperp^o}^{\kperpstar} \rmd\kperp \:  \frac{ E_\perp^+(\kperp)}{\noe \Toe}\right], 
	\label{eq:gamma_eff_definition}
\end{align}
which provides a proxy for the perpendicular-wavenumber-dependent damping rate \cref{eq:damping_rate_final}. Our theory predicts that $\gammaeff$ also has the same scaling with $\beta_e$, viz., $\gammaeff \propto 1/\sqrt{\beta_e}$.

\begin{figure*}
	\centering
	
	\begin{tikzonimage}[width=\figscale\textwidth]{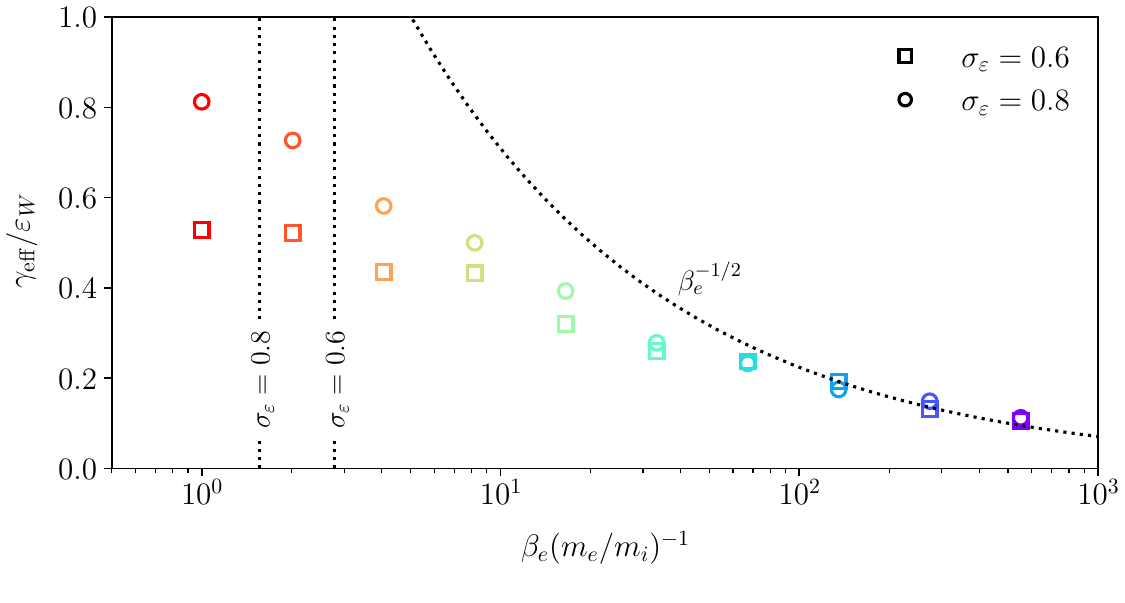}
		\node[scale=1.0] at (0.2645, 0.9) {\cref{eq:betacrit}};
	\end{tikzonimage}
	
	\caption[]{The effective damping rate \cref{eq:gamma_eff_definition} for the `beta scan' simulations in \cref{tab:simulation_parameters}, normalised to the total energy flux $\fluxe$, as a function of the normalised electron plasma beta $\beta_e (m_e/m_i)^{-1}$ (which is also indicated by the colours). The injection imbalances of $\forcing = 0.6, \: 0.8$ are shown by the circle and square markers, respectively. The vertical dotted lines indicate the values of the critical beta \cref{eq:betacrit} for each injection imbalance, while the curved dotted line shows the expected scaling $\gammaeff \propto \beta_e^{-1/2}$ at $\beta_e \gg \betacrit$.}  
	
	\label{fig:beta_scan_heating_rates}
\end{figure*}

To test this prediction, we consider the set of simulations labelled `beta scan' in \cref{tab:simulation_parameters}, which consists of otherwise-identical simulations at various values of $\beta_e$ logarithmically spaced in the range $1.0 \leqslant \beta_e (m_e/m_i)^{-1} \leqslant 550.9$, and for values of the injection imbalance of $\forcing = 0.6, \: 0.8$. In \cref{fig:beta_scan_heating_rates}, we plot the effective damping rate \cref{eq:gamma_eff_definition} for each of these simulations, averaged over times following the initial transient ($t/t_A \gtrsim 1$), and with $\kperpstar$ calculated as in \cref{sec:dependence_on_amplitude}. The simulations at $\beta_e \gg \betacrit$ reproduce the scaling expected from \cref{eq:damping_rate_final} well, while those at lower values depart from it. These deviations arise for two main reasons. First, \cref{eq:damping_rate_final} is derived from the KAW damping rate \cref{eq:gamma}, which applies only at scales $\kperp d_e \lesssim 1$ (see \cref{fig:dispersion_relation}). For $\beta_e \gg \betacrit$, this condition holds for the largest scales dominating the parallel-dissipation rate. However, as $\beta_e$ decreases, $d_e$ approaches $\rhoi$ from the small-scale side, and the linear damping rate deviates from \cref{eq:gamma}. In particular, the damping rate at scales $\kperp d_e \gtrsim 1$ is independent of beta, and so we would expect the scaling of the effective damping rate \cref{eq:gamma_eff_definition} to become weaker at lower values, which is manifest in \cref{fig:beta_scan_heating_rates}. A second, and arguably more important reason, for the discrepancies at lower beta is that the dynamics underlying the helicity barrier become less dominant: for $\beta_e \lesssim \betacrit$, the helicity barrier vanishes completely, with the system transitioning to the imbalance-steepened-cascade regime described in \cref{sec:weakened_cascade}, and in which \cref{eq:gamma_eff_definition} ceases to be applicable. Due to the dependence of $\betacrit$ on the injection imbalance [recall \cref{eq:betacrit}], this transition happens at a higher beta as $\forcing$ is decreased, manifest in \cref{fig:beta_scan_heating_rates}. While our results show that the helicity barrier is certainly absent for the lowest values of beta that we have considered [$\beta_e (m_e/m_i)^{-1} = 1.0$, the left-hand most point in \cref{fig:beta_scan_heating_rates}], what is not obvious is the nature of the transition between these two regimes, and indeed whether the critical beta \cref{eq:betacrit} needs to be modified for the fully kinetic KREHM system. This is the subject to which the next section is devoted. 

\subsection{Intermediate regime}
\label{sec:intermediate_regime}
In \cite{adkins24}, the critical beta \cref{eq:betacrit} arose as a result of the simultaneous conservation of both the (free) energy \cref{eq:free_energy} and (generalised) helicity \cref{eq:helicity} in the isothermal limit. The fact that these two nonlinear invariants were conserved at every perpendicular scale allowed the use of Kolmogorov-type arguments to obtain a condition for the existence of a constant-flux cascade, which in turn led to \cref{eq:betacrit}. Unfortunately, the same arguments cannot be repeated here because neither the energy flux nor the helicity flux remains constant: the former due to electron Landau damping (responsible for the cascade steepening discussed in \cref{sec:weakened_cascade}), the latter due to the presence of the source/sink of helicity \cref{eq:helicity_injection}. Nevertheless, it appears that the transition between the imbalanced-steepened-cascade and helicity-barrier regimes is still controlled by the critical beta \cref{eq:betacrit}, even if the transition that it demarcates is less `sharp' than in the isothermal case.\footnote{In the simulations considered by \cite{adkins24}, useful measures of the turbulence were completely distinct on either side of \cref{eq:betacrit}, irrespective of the value of the injection imbalance $\sigma_\varepsilon$ (see, e.g., their figures 2 and 3).}

\begin{figure*}
	\centering
	\includegraphics[width=\figscale\textwidth]{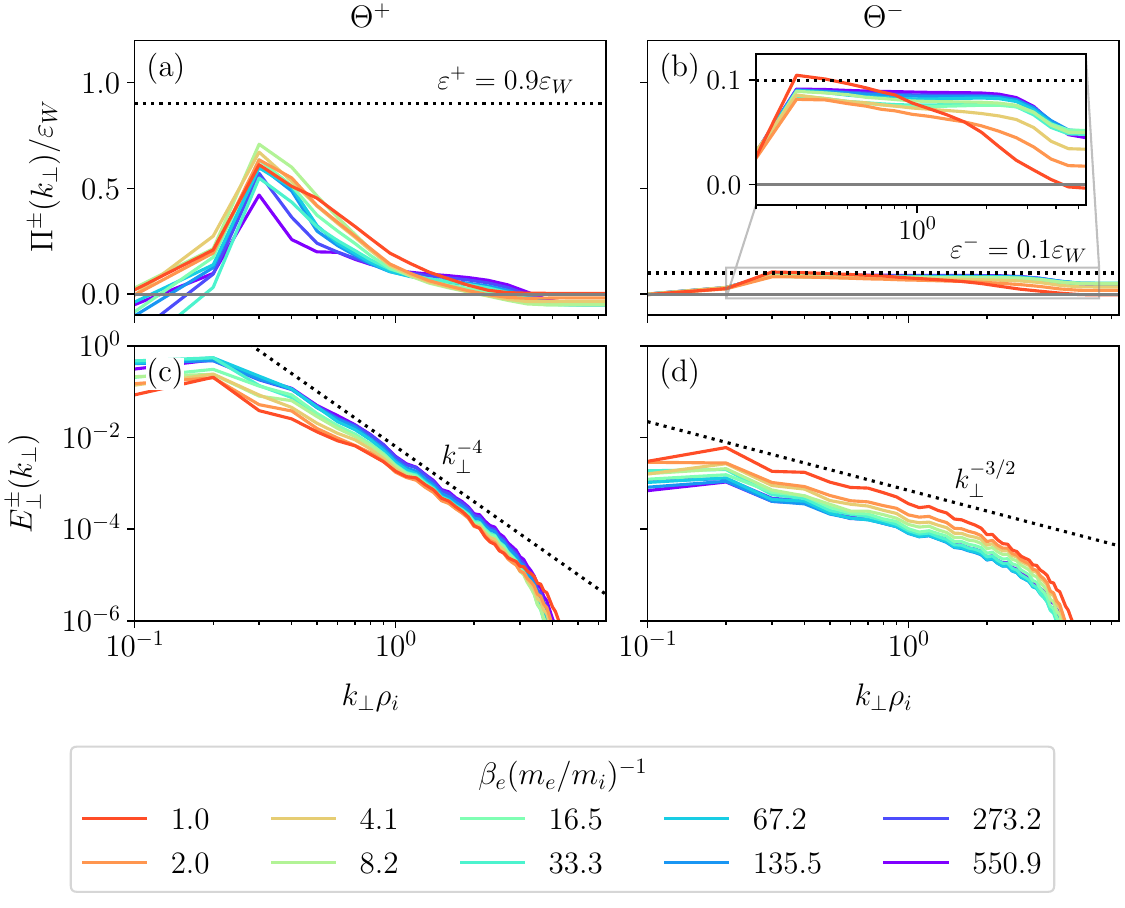}
	
	\caption[]{Time-averaged, one-dimensional perpendicular energy fluxes and spectra for the $\forcing = 0.8$ `beta scan' simulations in \cref{tab:simulation_parameters}, with the left- and right-hand columns showing results for the $\thetap$ and $\thetam$ fields, respectively. (a),(b) Energy fluxes $\Pi^\pm(\kperp)$ computed directly from the nonlinear terms in \cref{eq:phi_equation} and \cref{eq:apar_equation} and normalised to the total energy flux $\fluxe$. The horizontal dotted lines indicate the values of the fluxes \cref{eq:fluxpm} expected if the system were to maintain a constant-flux cascade. (c),(d) Perpendicular energy spectra $E_\perp^\pm(\kperp)$, defined in \cref{eq:thetapm_spectra_def}. The colour indicate the different values of $\beta_e(m_e/m_i)^{-1}$. The dotted lines show spectral scalings as labelled.}
	\label{fig:beta_scan_spectra}
\end{figure*}

\begin{figure*}
	\centering
	
	\begin{tikzonimage}[width=\figscale\textwidth]{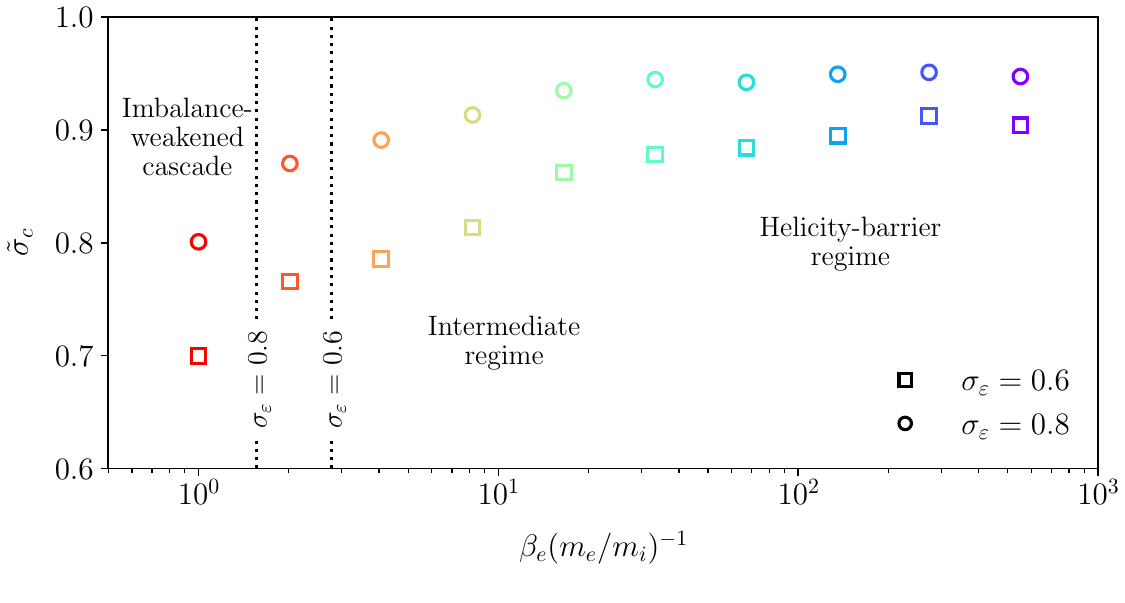}
		\node[scale=1.0] at (0.2625, 0.9) {\cref{eq:betacrit}};
	\end{tikzonimage}
	
	\caption[]{The energy imbalance $\imbalance = H/W$ for the `beta scan' simulations in \cref{tab:simulation_parameters}, averaged over the last 20\% of the simulation time, as a function of the normalised electron plasma beta $\beta_e(m_e/m_i)^{-1}$ (which is also indicated by the colours). The injection imbalances of $\forcing = 0.6, \: 0.8$ are shown by the circle and square markers, respectively. The vertical dotted lines indicate the values of the critical beta \cref{eq:betacrit} for each injection imbalance.}  
	
	\label{fig:beta_scan_imbalance}
\end{figure*}   

To see this, we plot, in \cref{fig:beta_scan_spectra}, the one-dimensional perpendicular energy fluxes and spectra for the $\forcing = 0.8$ `beta scan' simulations in \cref{tab:simulation_parameters}, for which $\betacrit (m_e/m_i)^{-1} = 1.56$. Let us first consider the simulation with $\beta_e (m_e/m_i)^{-1} = 550.9$ (purple curves), which, given the results of \cref{sec:barrier_regime}, certainly lies in the helicity-barrier regime. While the $\thetam$ flux [panel (b)] is nearly perfectly constant as a function of $\kperp$, the $\thetap$ flux [panel (a)] displays a sharp decrease at large scales before becoming approximately constant at small scales. This behaviour is typical of the helicity barrier \citep{meyrand21,squire22,squire23,adkins24,johnston25}, which only allows the balanced portion of the injected energy flux ($\approx 2\fluxm$) to cascade past the scale of the break, which here lies at $\kperpstar \rhoi \approx 0.7$. On the other hand, the simulation with $\beta_e (m_e/m_i)^{-1} = 1.0$ (red curves) shows a decrease in \textit{both} the $\thetap$ and $\thetam$ fluxes that continues to even the smallest scales in the simulation, behaviour that is well described by the imbalance-steepened cascade theory of \cref{sec:weakened_cascade}. As $\beta_e$ is increased from this value, the $\thetap$ flux gradually flattens at small scales, with the $\thetam$ flux instead flattening \textit{throughout} the inertial range. For $\beta_e (m_e/m_i)^{-1} \gtrsim 16.5$ (neon green), both fluxes have behaviour very similar to those in the $\beta_e (m_e/m_i)^{-1} = 550.9$ simulation. This means that for $1.0 \lesssim \beta_e (m_e/m_i)^{-1} \lesssim 16.5$ there is some `intermediate' regime in which the dynamics lie on some continuum between the imbalance-weakened cascade (\cref{sec:weakened_cascade}) and helicity-barrier (\cref{sec:barrier_regime}) regimes. Such a transition is reflected in the increase of the energy imbalance $\imbalance = H/W$ with $\beta_e$ that is manifest in \cref{fig:beta_scan_imbalance}. In the presence of the helicity barrier, the growth of the large-scale $\thetap$ amplitudes leads directly to an increase in $\imbalance$, which continues until electron Landau damping becomes large enough to saturate the barrier. This lower Landau damping means that we would expect simulations with $\beta_e \gg \betacrit$ to have higher energy imbalances (at a given time) than those at $\beta_e \ll \betacrit$, which is indeed what we observe. The finite width of the intermediate regime (also seen for $\forcing = 0.6$) arises as a result of the presence of electron kinetics disrupting the robust helicity-barrier formation observed in the isothermal limit --- either by damping the barrier directly or disrupting the conservation of generalised helicity \cref{eq:helicity}. In particular, we expect that at a given time (or at saturation if another method of parallel dissipation is present), the imbalance will increase with $\beta_e$ until the point where electron Landau damping is effectively negligible, after which the imbalance will become approximately independent of beta. \Cref{fig:beta_scan_imbalance} is tentative evidence that the `effective' critical beta in the presence of electron Landau damping occurs at a higher $\beta_e$ than the theoretical prediction \cref{eq:betacrit}, though more simulations at a variety of injection imbalances would be required to confirm this.

A final feature to note is that the $\thetap$ spectrum [panel (c) in \cref{fig:beta_scan_spectra}] shows a steep $\sim \kperp^{-4}$ scaling for all values of $\beta_e(m_e/m_i)^{-1}$, as we saw previously in \cref{fig:weakened_cascade_imbalanced}. Of course, the extent of the steepening observed in the $\thetap$ spectrum in the imbalanced-weakened cascade regime will itself depend on the (injection) imbalance [recall \cref{eq:fluxm_kperp} and \cref{eq:fluxp_kperp}], meaning that the difference in the spectrum between the imbalanced-weakened cascade and helicity-barrier regimes should be more obvious at lower imbalances. This has the interesting implication, already noted in \cref{sec:weakened_cascade}, that steep `transition range' spectra should be observed in the highly-imbalanced solar wind irrespective of the value of $\beta_e$ and, therefore, the presence, or otherwise, of the helicity barrier. We discuss this further in \cref{sec:summary}. 

\section{Consequences for solar-wind heating}
\label{sec:consequences_for_sw_heating}
Our theory, supported by numerical simulations, predicts that the nature of the turbulence, and thus the resultant heating, is different depending on whether the system lies above or below the critical beta \cref{eq:betacrit}. For $\beta_e \lesssim \betacrit$ --- the imbalanced-steepened cascade regime, \cref{sec:weakened_cascade} --- the presence of electron Landau damping reduces the flux of energy arriving to the smallest perpendicular scales, leading to a steepening of the associated perpendicular energy spectra that becomes more pronounced with increasing imbalance. For $\beta_e \gtrsim \betacrit$ --- the helicity-barrier regime, \cref{sec:barrier_regime} --- a helicity barrier forms, allowing only the balanced portion of the energy flux to continue its cascade past the spectral break at $\kperpstar$. The electron Landau damping is then dominated by scales $\kperp \lesssim \kperpstar$ due to the steep $\sim \kperp^{-4}$ `transition range' spectra below the break, which results in a specific form of the parallel-dissipation rate \cref{eq:parallel_diss_integral}. The intermediate regime (\cref{sec:intermediate_regime}) represents a gradual shift between the imbalanced-steepened cascade and helicity-barrier regimes, where neither process is fully dominant. As $\beta_e$ increases past $\betacrit$, the turbulence transitions from being strongly affected by electron Landau damping at all perpendicular scales to a state where a helicity barrier begins to form but is still partially disrupted by kinetic effects. This transition occurs over a finite range of $\beta_e$, rather than at a sharp threshold as was the case in the isothermal limit \citep{adkins24}. Because significant spectral steepening occurs in both regimes under different conditions, distinguishing between the imbalance-weakened cascade and helicity-barrier regimes based on spectra alone can be challenging (see \cref{fig:weakened_cascade_imbalanced} or \cref{fig:beta_scan_spectra}), regardless of the presence, or otherwise, of an intermediate regime.

We have, however, not yet engaged with the question of what will be the saturated amplitudes and electron-ion energy partition obtained by the turbulence in either of these regimes. This is due, in part, to the fact that any attempt to address this question solely within the KREHM framework is ultimately insufficient for application to real physical systems due to the absence of non-trivial ion physics within the model. In particular, there is no route by which any of the injected energy flux can be dissipated on ions. This omission is particularly significant given that ions are observed to be hotter than electrons in the solar wind, which implies that the ratio of the electron to ion heating rates, $Q_e/Q_i$, must be less than unity for at least some portion of its outward propagation \citep{cranmer09,bandyopadhyay23}. This is further complicated by the fact that in the helicity-barrier regime, the saturated amplitude reached by the turbulence is determined entirely by the mechanisms of dissipation at small \textit{parallel} scales \citep[see, e.g.,][]{meyrand21,squire22,squire23,adkins24}, which, as discussed in \cref{sec:barrier_regime}, can lie outside the gyrokinetic approximation. In lieu of attempting to predict both the electron-ion heating ratio and this saturated amplitude simultaneously, we assume that the system has reached some steady state with a given saturated amplitude, from which we then develop a theory for $Q_e/Q_i$ by synthesising insights from our study of KREHM with those from previous investigations of ion heating. Given the assumptions under which KREHM is derived, we restrict our considerations to plasmas with $\beta_e \lesssim 1$.  

\subsection{Heating channels}
\label{sec:heating_channels}
Steady-state energy conservation requires that the total injected energy flux must be balanced by the total heating rate, which can be expressed as a sum of the contributions from each plasma species, viz.,
\begin{align}
	\fluxnorm \fluxe = \qperpe + \qparae + \qperpi + \qparai,
	\label{eq:energy_conservation}
\end{align}
where $Q_{\perp \s}$ and $Q_{\parallel \s}$ are, respectively, the parallel and perpendicular heating rates of species $\s$. Note that in the following discussion we will ignore the contribution to the heat from minor-ion species (e.g., $\text{He}^{2+}$), since their dilute nature suggests that they will make at most a modest contribution. Let us now briefly review the physical mechanisms that may contribute to each of the terms in \cref{eq:energy_conservation}, providing context for our discussion of the imbalance-steepened cascade and helicity-barrier regimes that follows in \cref{sec:heating_imbalance_steepened} and \cref{sec:heating_helicity_barrier}, respectively. The interested reader may find a more comprehensive review of these mechanisms for balanced turbulence in \cite{howes24}.

\subsubsection{Perpendicular electron heating}
\label{sec:perpendicular_electron_heating}
Perpendicular electron heating $\qperpe$ occurs due to small-scale dissipation resulting from perpendicular (finite-Larmor-radius induced) phase mixing that occurs on scales $\kperp \rhoe \gtrsim 1$ \citep{sch09}, as well as the viscosity or resistivity arising from finite electron collisionality. While we formally neglected collisions within our analysis, the cascade of energy to fine (perpendicular) spatial scales means that collisions will inevitably always be activated, no matter how small their collision frequency is taken to be. While viscosity typically becomes important on scales ${\kperp \rhoe \gtrsim 1}$, resistivity may become dynamically important on scales larger than this. We can estimate the perpendicular scale at which this occurs by balancing the linear frequency and the resistive rate, viz., $\omega \sim \kperp^2 d_e^2 \nu_{ei}$, which, approximating $\omega \sim \kpar \vphase$ by the isothermal result \cref{eq:linear_frequency} and assuming that $\tau$ is not small, becomes \citep[cf.][]{sch09}:
\begin{align}
	\kperp \rhoe \sim \kpar \mfpi \sqrt{\beta_e} \left(\frac{Z}{\tau}\right)^{3/2},
	\label{eq:resistive_scale}
\end{align}
where $\mfpi = \vthi/\nu_{ii}$ is the mean-free-path for ion-ion collisions, and $\nu_{\s\s'}$ is the collision frequency between species $\s$ and $\s'$.
Since the parallel scales of the turbulence are always comparable to or smaller than those at the outer scale, the right-hand side of \cref{eq:resistive_scale} is bounded from below by the value of $\kpar \mfpi$ at the outer scale. For a value of $\kpar^o \mfpi = 10^3$, typical of the solar wind, the right-hand side of \cref{eq:resistive_scale} is greater than unity for even the extreme values of $\beta_e = m_e/m_i$ and $\tau = 10$. This implies that perpendicular electron heating, whether through viscosity or resistivity, is confined to scales $\kperp \rhoe \gtrsim 1$ for $\beta_e \lesssim 1$. While this conclusion may seem obvious, it has the important implication that the energy flux available to be dissipated by $\qperpe$ is that which remains after all of the other mechanisms of dissipation are accounted for, which are each active on larger perpendicular scales.

\subsubsection{Parallel electron heating}
\label{sec:parallel_electron_heating}
Parallel electron heating $\qparae$ can, in principle, arise from two distinct mechanisms. The first is so-called `transit-time damping' \citep{barnes66}, which relies on the thermal pressure being sufficiently large that it can excite compressive magnetic-field perturbations that are subsequently damped on mirror-resonant electrons. However, in the $\beta_e \lesssim 1$ limit considered here, the magnetic pressure dominates over the thermal one, suppressing such perturbations and rendering the effects of transit-time damping largely negligible. The only remaining mechanism for parallel electron heating is then electron Landau damping, which has already been discussed at length in this paper. As such, we will not dwell on it further here, except to emphasise the following key point. 
In balanced turbulence, significant electron Landau damping is typically, for $\beta_e \gtrsim m_e/m_i$, localised on scales $\kperp \rhoi \gtrsim 1$ \citep{howes08jgr,howes11cascade,howes24} due to the perpendicular-wavenumber dependence of the linear damping rate (see \cref{sec:linear_dispersion_relation}) meaning that it is vanishingly small on larger scales. In imbalanced turbulence, however, we have seen that significant electron Landau damping can occur even on scales $\kperp \rhoi \lesssim 1$, increasing the fraction of the injected energy flux that is, in principle, able to contribute to parallel electron heating. This distinction will be key for the considerations that follow. 

\subsubsection{Perpendicular ion heating}
\label{sec:perpendicular_ion_heating}
Two widely studied mechanisms of perpendicular ion heating $\qperpi$ in the solar wind are stochastic heating and quasilinear cyclotron resonant diffusion. In the former, ions are heated by random/uncorrelated deflections from turbulent fluctuations on scales comparable to the ion Larmor radius, leading to the diffusive heating of the ion kinetic distribution function in perpendicular velocities \citep{mcchesney97,johnson01,chandran10,xia13,hoppock18,arzamasskiy19,cerri21}. In the latter, resonant interactions between the ions and waves around the ion-Larmor frequency $\Omega_i$ cause ions to undergo quasilinear diffusion along contours of constant energy in the frame moving at the phase velocity of the waves \citep{kennel66,stix92,schlickeiser93}, heating the plasma. While these have traditionally been considered as separate mechanisms of ion heating, recent work by \cite{johnston25} argues that both stochastic and quasilinear heating can be thought of as two limits of some continuum controlled by the broadening of the fluctuations' frequency spectrum. Unbroadened fluctuations well-approximated by a single frequency $\omega$ and wavenumber $\vk$ cause quaslinear-heating-like behaviour, while fluctuations that are broadened to a level comparable to $\omega$ itself cause stochastic-heating-like behaviour (having more power in $\omega < \tnl^{-1}$ fluctuations). \cite{johnston25} argue that the broadening is directly related to imbalance: imbalanced turbulence is energetically dominated by $\thetap$ fluctuations that have linear frequencies exceeding their nonlinear rate (the quasilinear-heating limit), while the two are comparable in balanced turbulence (the stochastic-heating limit). Thus, in what follows, we will consider perpendicular ion heating as the unified effect of both mechanisms, rather than treating them as distinct processes. Importantly, assuming critical balance, both mechanisms require sufficiently high amplitudes in order to contribute meaningfully to the energy balance in \cref{eq:energy_conservation}. \cite{chandran10,xia13} found that a minimum amplitude for the velocity fluctuations at scales comparable to the ion Larmor radius of $\delta u_{\rhoi}/\vthi \sim 1$ was required for the onset of stochastic heating, and \cite{johnston25} argued that a very similar threshold also applies in the quasilinear limit.

\subsubsection{Parallel ion heating}
\label{sec:parallel_ion_heating}
In a similar vein to electrons, parallel ion heating $\qparai$ arises from either ion transit-time damping or ion Landau damping. The former is once again weak at $\beta_e \lesssim 1$ for the same reasons as those discussed in \cref{sec:parallel_electron_heating}. However, unlike in the electron case, ion Landau damping is also weak in this regime for the same reason that the ion kinetics were neglected within KREHM: the typical frequencies of the (Aflv\'enic) perturbations are significantly faster than the parallel-streaming rate of the ions. This means that the Landau resonance responsible for the damping occurs far out in the `tail' of the ion distribution function ($\omega/ \kpar \sim \vphase \gg \vthi$), where both its gradient and the associated number of particles are small, preventing ions from efficiently resonating with the turbulent fluctuations and suppressing the associated rate of damping. As a result, parallel ion heating is expected to be dynamically unimportant except at sufficiently high $\beta_i \sim 1$ (where $\omega /\kpar \sim  \vthi$) and so we will henceforth neglect it from our analysis.

\subsection{Imbalance-steepened cascade regime}
\label{sec:heating_imbalance_steepened}
Let us first consider the $\beta_e \lesssim \betacrit$ regime, in which the cascade of energy to small perpendicular scales is not inhibited by the presence of a helicity barrier. Previous studies of balanced gyrokinetic turbulence have shown that electron heating significantly exceeds ion heating when the cascade is predominantly Alfv\'enic in character, i.e. negligible energy is contained in the compressive fluctuations \citep[][]{howes08jgr,howes10,howes11cascade,kawazura19,howes24}. We note, however, that it is possible to obtain significant perpendicular ion heating (via either of the mechanisms discussed in \cref{sec:perpendicular_ion_heating}) at larger amplitudes that lie outside the gyrokinetic approximation \citep[see, e.g.,][]{arzamasskiy19,cerri21}. Now, the results of \cref{sec:weakened_cascade} indicate that the presence of non-zero imbalance enhances the overall dissipation of the energy flux relative to the balanced case when considering the electron dynamics in isolation [note, in particular, the dependence of the $\thetap$ flux \cref{eq:fluxp_kperp} on the imbalance]. This enhanced dissipation steepens the perpendicular energy spectrum (see \cref{fig:weakened_cascade_imbalanced}) on scales $\kperp \rhoi \sim 1$ where ion heating is usually most significant. Consequently, the turbulent amplitudes at these scales are reduced compared to the balanced case, which would, presumably, further weaken the ion heating. Thus, we expect electron heating to remain dominant in the imbalance-steepened cascade regime, except when driven extremely strongly. Mathematically, this can be expressed as
\begin{align}
	\frac{Q_e}{Q_i} = G(\sigma_c, \beta_e, \tau, Z, \kperp^o \rhoi, \: \dots) \gg 1,
	\label{eq:heating_imbalance_steepened}
\end{align}
for some function $G$ that depends on the parameters of turbulence including, but not limited to, the normalised cross-helicity, electron beta, temperature ratio, charge, and outer scale, listed here from left to right. This function could be evaluated explicitly through analogous methods to those used in \cite{howes08jgr,howes24} or \cite{chandran10,xia13} --- taking into account the effect of non-zero imbalance by modifying the perpendicular energy fluxes according to the prescription of \cref{sec:elsasser_fluxes} --- though to do so here would be beyond the scope of the current study. This result is, in some sense, a natural consequence of continuity: imbalanced turbulence without a helicity barrier must transition smoothly to the balanced regime at lower levels of imbalance, and \cref{eq:heating_imbalance_steepened} reflects this continuity in the partitioning of heating between electrons and ions, combined with the expectation that the spectral steepening due to electron Landau damping should further reduce ion heating. The same is not true for the helicity-barrier regime, the subject of the following section.

\subsection{Helicity-barrier regime}
\label{sec:heating_helicity_barrier}

\subsubsection{Electron heating rate}
\label{sec:electron_heating_rate}
In the presence of the helicity barrier, the amount of injected energy flux that is allowed to cascade past the spectral break, i.e., to scales $\kperp \gtrsim\kperpstar$, is limited to the balanced portion $\approx 2\fluxm$. While the results of \cref{sec:barrier_regime} show that electron Landau damping is not entirely negligible on scales $\kperpstar \lesssim \kperp \lesssim \rhoe^{-1}$, its effects are most significant for the imbalanced portion of the energy at $\kperp \lesssim \kperpstar$ [see the dissipation spectra in \cref{fig:restart_spectra_perp}(b)]. This means that the balanced portion of the flux survives to the smallest scales, being roughly constant as a function of perpendicular wavenumber [see \cref{fig:beta_scan_spectra}(a),(b) for the highest beta simulations], and so we expect that the perpendicular electron heating rate will be well-approximated by\footnote{The appearence of the factor of $\fluxnorm$ in \cref{eq:electron_heating_perp_original} is due to our definition of $\fluxm$ as a rate, with units of inverse time [see, e.g., \cref{eq:free_energy_budget}].}
\begin{align}
	\qperpe \sim 2 \fluxnorm \fluxm.
	\label{eq:electron_heating_perp_original}
\end{align}
Given the constancy of $\fluxm$, we can express it straightforwardly in terms of the amplitudes at the outer scale $\kperp^o$ [note that this is not the case for $\fluxp$, which is a decreasing function of perpendicular wavenumber due to the helicity barrier; see \cref{fig:beta_scan_spectra}(a)]. Then, \cref{eq:electron_heating_perp_original} can be written as
\begin{align}
	\frac{\qperpe}{\noi m_i \vthi^2 \Omega_i} & \sim {2}(\kperp^o \rhoi) \left(\frac{\zedpsouter}{\vthi}\right) \left(\frac{\zedmsouter}{\vthi}\right)^2 \nonumber\\
	& \sim {2}(\kperp^o \rhoi) \left(\frac{1-\sigma_c}{1+\sigma_c}\right) \left(\frac{\zedpsouter}{\vthi}\right)^3,
	\label{eq:electron_heating_perp} 
\end{align}
where the final expression is obtained by using the definition of the normalised cross-helicity and \cref{eq:nonlinear_times_ratio} to relate the outer-scale amplitudes to the imbalance. In \cref{eq:electron_heating_perp}, $\zedpms \equiv \kperp \thetapms$ are the characteristic amplitudes of the `Els\"asser fields' corresponding to the generalised Els\"asser potentials \cref{eq:thetapm} [see \cref{eq:thetapm_spectra_def}], which reduce to their RMHD counterparts $z^\pm_{\kperp}$ on large scales $\kperp \rhoi \ll 1$. We will adopt this notation throughout the remainder of \cref{sec:consequences_for_sw_heating} in order to make the connection to the RMHD limit more explicit in expressions that follow.

To estimate the parallel electron heating rate $\qparae$, we return to our expression for the parallel-dissipation rate \cref{eq:parallel_diss_integral}. Recall that in deriving our expression \cref{eq:damping_rate_final} for the damping rate appearing therein, we assumed that the perpendicular energy spectrum of $\thetap$ followed $E_\perp^+ \propto \kperp^{-\alphap}$ on scales $\kperp \lesssim \kperpstar$. As discussed following \cref{eq:damping_rate_final}, for typical values of $\alphap$, this means that the integral in \cref{eq:parallel_diss_integral} will be dominated by its value at the upper bound $\kperpstar$. Assuming that there exists sufficient separation between this and the outer scale $\kperp^o$, as would be the case in a realistic plasma system, we can evaluate the integral directly, yielding:
\begin{align}
	\frac{D_\parallel}{\Omegai} \simeq & \:\frac{4(\alphap - 1)\cpar}{3(3 - \alphap)} \left(\frac{d_e}{\rhos}\right) 	\nonumber \\
	& \times \left(\kperp^o \rhos\right)^{3(\alphap-1)/2}\left(\kperpstar \rhos\right)^{(9-3\alphap)/2} \left(\frac{W^+}{\noe\Toe}\right)^{3/2}. \label{eq:parallel_diss_integrated}
\end{align}
From \cref{eq:free_energy_thetapm}, we can estimate the free energy of the $\thetap$ fluctuations appearing in \cref{eq:parallel_diss_integrated} by assuming that it is dominated by its outer-scale contribution, viz., 
\begin{align}
	\frac{W^+}{\noe \Toe} \sim \left(\frac{\kperp^o \Theta^+_{\kperp^o}}{c_s}\right)^2 \sim \left(\frac{\zedpsouter}{c_s}\right)^2.
	\label{eq:free_energy_outer}
\end{align}
Finally, using the fact that $\qparae \sim \noe \Toe D_\parallel$ [see \cref{eq:free_energy_budget}], we can write down an estimate for the electron heating rate:
\begin{align}
	\frac{\qparae}{\noi \mi \vthi^2 \Omegai} \sim &\frac{ Z^{3/2}  \constonee}{\tau} \left(\frac{m_e}{\beta_e m_i}\right)^{1/2} \nonumber\\
	&\times \left(\kperp^o \rhoi\right)^{3(\alphap-1)/2}\left(\kperpstar \rhoi\right)^{(9-3\alphap)/2}  \left(\frac{\zedpsouter}{\vthi}\right)^3,
	\label{eq:electron_heating_para}
\end{align}
in which $\constonee = 2^{3/2}(\alphap - 1)\cpar/(9 - 3 \alphap)$ is an order-unity constant. The form of \cref{eq:electron_heating_para} highlights the important role that imbalance plays in determining the magnitude of the parallel electron heating. For $\alphap < 3$, \cref{eq:electron_heating_para} contains a positive power of $\kperpstar$ which, by \cref{eq:kperpstar}, means that it is proportional to some positive power of $1 - \sigma_c$. The reason for this dependence is somewhat intuitive: higher imbalances causes the break to shift to larger scales where the linear Landau-damping rate is weaker [see \cref{eq:damping_rate_final}], thus decreasing $\qparae$. Additionally, this heating is more effective at lower beta due to the fact that the linear Landau damping rate reaches its maximum at $\kperp d_e \sim 1$ [see \cref{fig:dispersion_relation}(b)], which moves to larger scales as $\beta_e$ is decreased, although we emphasise that we must always have $\beta_e \gtrsim \betacrit$ in \cref{eq:electron_heating_para} in order for the prediction to be valid. 

\subsubsection{Ion heating rate}
\label{sec:ion_heating_rate}
Let us now turn our attention to the ion heating rate. As we have seen, the presence of the helicity barrier leads to the growth of turbulent amplitudes on scales $\kperp \lesssim \kperpstar$. Hybrid-kinetic simulations of the helicity barrier \citep{squire22,squire23,zhang25} have shown that by increasing the turbulent amplitudes it enhances the perpendicular ion heating, allowing it to play a significant role in the overall energy balance. To model this $\qperpi$, we adopt equation (1) of \cite{johnston25}:
\begin{align}
	\frac{\qperpi}{\noi \mi \vthi^2 \Omegai} =  \constonei  \xii^3 F(\xii, \dots),
	\label{eq:ion_heating_original} 
\end{align}
since it appears to capture ion heating across a wide variety of different turbulence properties, including in the presence of the helicity barrier. In \cref{eq:ion_heating_original}, $\constonei$ is some order-unity empirical constant \citep{chandran10,johnston25}, $\xii = (\kperp \rhoi)^{1/3} (\zedps/\vthi)$ is a normalised measure of the turbulent amplitudes, and $F(\xii, \dots)$ is some suppression factor that accounts for the (exponentially) small number of particles that are able to be resonant with the lowest-frequency fluctuations. In principle, $F(\xii, \dots)$ a function of other parameters of the system (e.g., imbalance), as well as $\xii$ itself, but we will not engage further with the details of this suppression factor here, focussing instead on a regime where the ion heating is allowed to play a significant role in the dynamics and for which $F(\xii, \dots ) \sim 1$. As in \cite{johnston25}\footnote{More generally, the $\zedps$ appearing in $\xii$ is the characteristic turbulent amplitude at $\kperp \rhoi \lesssim 1$ corresponding to the perpendicular scale $\kperp^{-1}$ at which the fluctuations reach their highest frequency, with the associated parallel scale $\kpar^{-1}$ determined from the critical-balance condition \cref{eq:critical_balance}. For example, in the balanced regime, this scale would be $\kperp \rhoi \sim 1$ \citep{johnston25}, for which \cref{eq:ion_heating_original} becomes the original stochastic-heating formula \citep{chandran10}. In the presence of the helicity barrier, however, this will occur at $\kperp = \kperpstar$, since, as was the case for the electron heating rate in \cref{sec:barrier_regime}, the existence of the sharp spectral break means that there is little energy available to be thermalised at $\kperp \gtrsim \kperpstar$.}, the scale $\kperp^{-1}$ in $\xii$ is taken to be the scale of the spectral break ${\kperpstar}^{-1}$ which, again assuming that the perpendicular spectrum of $\thetap$ follows $E_\perp^+ \propto \kperp^{-\alphap}$, allows us to write the perpendicular ion heating rate \cref{eq:ion_heating_original} as\footnote{The specific dependence of \cref{eq:ion_heating_rate} on $\kperpstar$ is a result of our implicit assumption of isotropic fluctuations in the perpendicular plane, for which $\alphap = 5/3$. This would have to be modified in the presence of dynamically-aligned and/or anisotropic fluctuations, though how to do so in the presence of a non-constant flux remains an open research question, and it is entirely possible that \cref{eq:ion_heating_rate} could still be true even in the presence of such fluctuations. To avoid these potential complications, we will use the isotropic result $\alphap=5/3$ whenever we explicitly evaluate \cref{eq:ion_heating_rate} and other expressions involving it.}
\begin{align}
	\frac{\qperpi}{\noi \mi \vthi^2 \Omegai} \sim \constonei \left(\kperp^o \rhoi\right)^{3(\alphap-1)/2}\left(\kperpstar \rhoi\right)^{(5-3\alphap)/2}  \left(\frac{\zedpsouter}{\vthi}\right)^3.
	\label{eq:ion_heating_rate}
\end{align}
Unlike in \cref{eq:electron_heating_para}, there is no dependence on $\beta_e$ due to the fact that \cref{eq:ion_heating_original} was derived without any knowledge of the electron dynamics. Finally, following the discussion of \cref{sec:parallel_ion_heating}, we assume that the parallel ion heating rate can be neglected when compared with the perpendicular one, such that $Q_i \approx \qperpi$ --- this is theoretically justified for quasilinear heating at $\beta_e \ll 1$, and is supported by hybrid-kinetic simulations of the helicity barrier \citep{squire22,squire23,zhang25}. 

\subsubsection{Electron-ion heating ratio}
\label{sec:electron_ion_heating_rate}
\begin{figure*}
	\centering
	
	\begin{tikzonimage}[width=\figscale\textwidth]{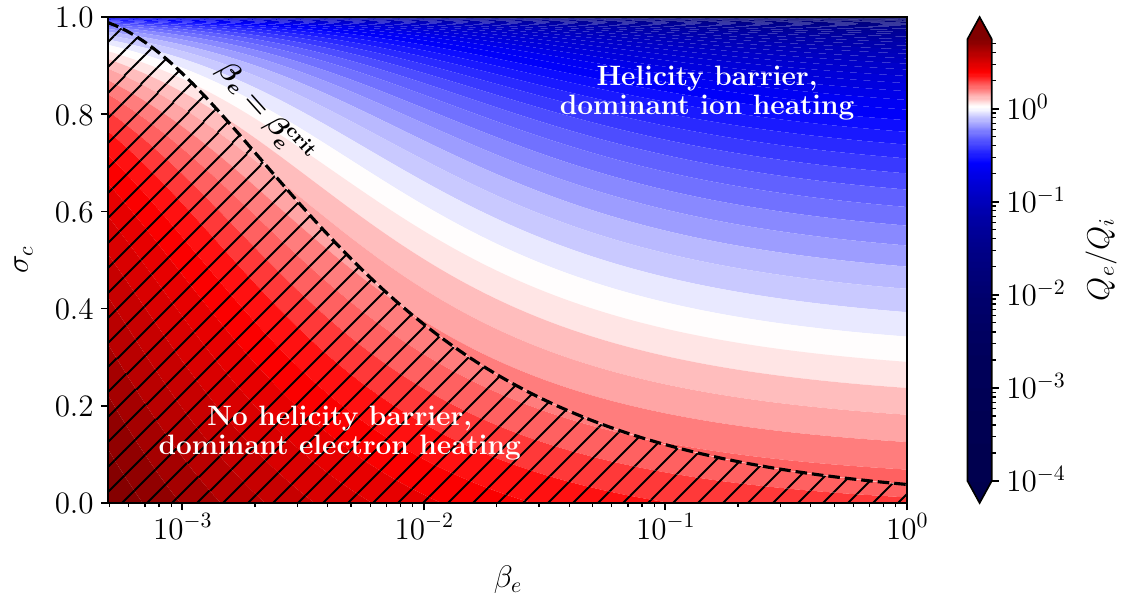}
	\end{tikzonimage}
	
	\caption[]{A contour plot of the electron-ion heating ratio $Q_e/Q_i$ \cref{eq:heating_helicity_barrier} as a function of the electron beta $\beta_e$ and normalised cross-helicity $\sigma_c$ for $Z =1$, $\tau = 2$, $\alphap = 5/3$, with the constants therein taking values $\constonei = 1.1$ \citep{johnston25}, $\constonee = 2.0$ (evaluated using $\cpar = 4.2$, the value from our simulations; see \cref{fig:restart_heating_rates} and the surrounding discussion), and $\constar = 2.0$ \citep{squire23}. The hatching indicates the region below the critical electron beta \cref{eq:betacrit} (black dashed line, evaluated using the results of \citealt{lithwick07}) in which \cref{eq:heating_helicity_barrier} is no longer valid, and the electron-ion heating ratio must be evaluated by other methods (see \cref{sec:heating_imbalance_steepened}).}  
	
	\label{fig:heating_ratio}
\end{figure*} 

Summing the parallel and perpendicular electron heating rates [\cref{eq:electron_heating_perp} and \cref{eq:electron_heating_para}, respectively], dividing the result by the ion heating rate \cref{eq:ion_heating_rate}, and using 
	\begin{align}
		\kperpstar \rhoi = \constar (1 - \sigma_c)^{1/4},
		\label{eq:kperpstar_const}
	\end{align}
	for some constant $\constar$, we obtain an expression for the electron-ion heating ratio in the presence of the helicity barrier:
	\begin{align}
		\frac{Q_e}{Q_i} = &\frac{ \constonee \constar^2}{\constonei}\frac{Z^{3/2}}{\tau} \left(\frac{m_e}{\beta_e m_i}\right)^{1/2} \left(1- \sigma_c\right)^{1/2} \nonumber \\
		&+ \frac{2}{\constonei} \left(\frac{\kperp^o \rhoi}{\constar}\right)^{(5-3\alphap)/2} \frac{\left(1 - \sigma_c\right)^{1 - (5-3\alphap)/8}}{1+\sigma_c}. 
		\label{eq:heating_helicity_barrier}
	\end{align}
The first and second terms in \cref{eq:heating_helicity_barrier} correspond to the contributions from parallel and perpendicular electron heating, respectively. For $\alphap > 1/3$, the first has a weaker dependence on $1-\sigma_c$ than the second, meaning that we expect  the parallel contribution to dominate over the perpendicular one at higher imbalance, even if the overall electron heating rate remains subdominant to the ion one. Conversely, the perpendicular component will dominate over the parallel one at higher $\beta_e$ due to the fact that the electron Landau damping rate becomes vanishingly small in this regime, while the flux of energy to small scales remains approximately constant.

It is important to note that \cref{eq:heating_helicity_barrier} cannot be considered a complete prediction for $Q_e/Q_i$ because, unlike in the imbalance-steepened cascade case, the normalised cross-helicity is itself implicitly a function of both $Q_e$ and $Q_i$. However, it can still be used to make nontrivial predictions about the relative importance of the electron and ion heating rates. In particular, an important implication of \cref{eq:heating_helicity_barrier} is that, in the saturated state, ion heating will generally dominate over electron heating in the helicity-barrier regime. The requirement for helicity-barrier formation set by the critical beta \cref{eq:betacrit} is $m_e/m_i \lesssim \betacrit \lesssim \beta_e$. Given that $\sigma_c \geqslant 0 $, $\kperp^o \rhoi \ll 1$, and $\tau \gtrsim 1$ in observations of the imbalanced solar wind, the contribution from the first term in \cref{eq:heating_helicity_barrier} will always be negligible in comparison to the second for any plasma with a value of $\beta_e$ sufficiently high to form a helicity barrier. This, coupled with the strong dependence of the latter on imbalance, means that we find $Q_e \lesssim Q_i$ at even modest imbalances. To illustrate this, we plot \cref{eq:heating_helicity_barrier} as a function of $\beta_e$ and $\sigma_c$ for a set of representative parameters in \cref{fig:heating_ratio}. It is immediately obvious that ion heating dominates over the vast majority of the parameter space where the helicity barrier operates, becoming comparable to the electron heating rate only for $\beta_e \approx \betacrit$. This provides \textit{post-hoc} justification for the use of the isothermal electron approximation in the imbalanced hybrid-kinetic simulations of \cite{squire22,squire23,zhang25} --- with all of the cases therein having $\beta_e \geqslant 0.3$ and $\sigma_c \geqslant 0.6$ in the saturated state --- and is consistent with solar-wind observations \citep{cranmer09,bandyopadhyay23}. We have chosen to only plot \cref{eq:heating_helicity_barrier}	in \cref{fig:heating_ratio} because creating a complete plot of $Q_e/Q_i$ valid across the whole parameter space would not only require a better understanding of the transition between the imbalance-steepened cascade and helicity-barrier regimes, but also evaluation of the right-hand side of \cref{eq:heating_imbalance_steepened}, both of which are beyond the scope of the current study.

Together, the combination of \cref{eq:heating_imbalance_steepened} and \cref{eq:heating_helicity_barrier} paint what could be described as a `winner takes all' picture of solar-wind heating: depending on the values of the electron beta and the imbalance, it is either entirely dominated by the electron channel, at lower beta and/or imbalance, or the ion one, at higher beta and/or imbalance. There are, of course, a number of important simplifications that were made in arriving at \cref{eq:heating_imbalance_steepened} and \cref{eq:heating_helicity_barrier} that could change this conclusion. The most straightforward of these is our neglect of the suppression factor in the ion heating rate \cref{eq:ion_heating_original} which could significantly decrease it relative to the electron one, modifying \cref{eq:heating_helicity_barrier} by a more-than-order-unity amount. Furthermore, careful readers will have noticed that it is possible for both of the predicted parallel electron and perpendicular ion heating rates [\cref{eq:electron_heating_para} and \cref{eq:ion_heating_rate}, respectively] to exceed the injected energy flux $\fluxnorm \fluxe \propto \zedmsouter (\zedpsouter)^2$, meaning that the system will have to adjust in ways not accounted for here to ensure that energy remains conserved. For example, restoring the suppression factor in \cref{eq:ion_heating_rate} would mean that a small decrease in the turbulent amplitudes would lead to a proportionally larger decrease in $\qperpi$ \citep{chandran10,xia13}. More generally, as stated above, the normalised cross-helicity $\sigma_c$ will itself be determined by $Q_e$ and $Q_i$, meaning that these individual rates may have to be modified when simultaneously present in a turbulent system. Unfortunately, the exact manner in which this has to be done will have to be informed by simulations involving both kinetic species, a task that lies outside the scope of this paper. Nevertheless, we expect that the scaling of \cref{eq:heating_helicity_barrier} with amplitude (implicit in its dependence on $\sigma_c$) will remain robust in the presence of such complexities. Specifically, for a fixed minus field, any increase in the amplitude of the energetically dominant plus field will act to decrease the relative electron heating rate while simultaneously enhancing the ion one, thereby reinforcing the distinct separation between electron- and ion-dominated heating regimes.

\section{Summary and discussion}
\label{sec:summary}
The findings of this paper demonstrate the crucial role of imbalance in determining the available dissipation channels and resultant heating in turbulent kinetic plasma systems. Using the KREHM system of equations, derived in the low-beta asymptotic limit of gyrokinetics, we showed that the critical beta \citep{adkins24} 
\begin{align}
	\betacrit = \frac{2Z}{1+\tau/Z} \frac{m_e}{m_i} \frac{1}{\forcing^2},
	\label{eq:betacrit_summary}
\end{align} 
remains the boundary between two fundamentally different regimes of turbulence and heating even in the presence of electron kinetics. Below this threshold, electron Landau damping reduces the otherwise constant flux of the injected free energy from large to small perpendicular scales, resulting in a steepening of the electromagnetic energy spectra. This steepening becomes significantly more pronounced at higher imbalances [see \cref{eq:fluxm_kperp}, \cref{eq:fluxp_kperp}, and the following discussion], and can even exhibit steep slopes on scales comparable to the ion Larmor radius [see \cref{fig:weakened_cascade_imbalanced}, panels (e),(f)]. This steepening can be distinguished the typical steepening associated with the `transition range' because it will not display the spectral flattening on smaller scales that usually accompany the latter. This is the regime of dominant electron heating, where the majority of the free energy injected on the largest perpendicular scales is dissipated on electrons in a manner similar to previous theories of low-beta, Alfv\'enic turbulence in the balanced regime \citep[see, e.g.,][]{howes08jgr,sch09,kawazura19,sch19}. Systems with $\beta_e$ above \cref{eq:betacrit_summary}, however, cannot sustain this `imbalance-steepened cascade'. Instead, a helicity barrier forms (\cref{sec:barrier_regime}), preventing all but the balanced portion of the injected free energy from cascading past the ion Larmor scale $\kperp \rhoi \sim 1$ and resulting in a significant increase in the turbulent amplitudes at larger scales $\kperp \rhoi \lesssim 1$. As in previous work \citep{meyrand21,squire23}, the location of the break in perpendicular-wavenumber space $\kperpstar$ is shown numerically to scale as $\kperpstar \rhoi \sim (1-\sigma_c)^{1/4}$ [see inset of \cref{fig:restart_spectra_perp}(b)], and we provide a novel theoretical justification of this scaling in appendix \ref{app:kperpstar}. The existence of the associated steep spectral slopes at $\kperp \gtrsim \kperpstar$ means that the parallel electron heating rate due to Landau damping is dominated by scales $\kperp \lesssim \kperpstar$, allowing us to derive an analytical expression for this heating rate that shows good agreement with numerical simulations (see \cref{fig:restart_heating_rates}). Predicting the electron-ion heating ratio $Q_e/Q_i$ \cref{eq:heating_helicity_barrier} by combining our results with those from previous investigations of ion heating reveals that the helicity-barrier regime remains one of dominant ion heating even in the presence of electron dissipation at these largest scales. This is because the electron heating rate has a stronger dependence on $\kperpstar \rhoi$ than the ion one [cf. \cref{eq:electron_heating_para} and \cref{eq:ion_heating_rate}], and so its relative heating rate will decrease faster as the imbalance and/or outer-scale amplitudes are increased. 

Taken together, our results appear to suggest a `winner takes all' picture of kinetic plasma heating in low-beta imbalanced Alfv\'enic turbulence --- the heating is dominated by either ions or electrons depending on the characteristics of the plasma at large scales, here captured through the electron plasma beta $\beta_e$ and the normalised cross-helicity $\sigma_c$. Such a picture has clear implications for observations. Given that much of the solar wind typically has $m_e/m_i \ll \beta_e \lesssim 1$ \citep{bruno05}, we would expect these plasmas to display clear features of helicity-barrier-mediated turbulence. This is indeed the case \citep{mcintyre24}: ions are hotter than electrons \citep{cranmer09,bandyopadhyay23}; there is significant power in ICWs around $\kpar \rhoi \sim 1$ \citep{huang20,bowen20waves,bowen24} due, in this paradigm, to quasilinear ion heating \citep[see][]{squire22,squire23,zhang25,johnston25}; and electromagnetic spectra observed by PSP usually show a steep `transition range' scaling $\sim \kperp^{-4}$ around $\kperp \rhoi \sim 1$ that has been seen in all simulations of the helicity barrier \citep{meyrand21,squire22,squire23,adkins24,zhang25,johnston25}. Our work also has interesting implications for observations of plasmas that are highly imbalanced but are at sufficiently low beta that the helicity barrier is not expected to form ($\beta_e \lesssim \betacrit$), such as in the low corona. In particular, the fact that the steepening of the electromagnetic fields due to electron Landau damping becomes more pronounced with increasing imbalance means that it should be possible to observe very steep spectra in the absence of other features of the helicity barrier, such as significant power in ICWs. 

There are, of course, a number of important questions left unanswered by this work. The arguments presented in \cite{meyrand21} and \cite{adkins24} for the breakdown of the constant-flux cascade solution, and the resultant formation of the helicity barrier, relied on the simultaneous conservation of both free energy \cref{eq:free_energy} and generalised helicity \cref{eq:helicity} throughout the inertial range; neither of these invariants are conserved in the presence of electron kinetics (see \cref{sec:nonlinear_invariants}). While our model based only on Landau damping provides a good fit to numerical results, we have not addressed the role played by the non-conservation of the generalised helicity in this context. In particular, this non-conservation of helicity may play a central role in the finite width in $\beta_e$ of the `intermediate regime' (see \cref{sec:intermediate_regime}) between the imbalance-weakened cascade and helicity-barrier ones. Understanding what sets this transition width --- and whether the effective shift of the critical beta \cref{eq:betacrit} towards higher $\beta_e$ seen in \cref{fig:beta_scan_imbalance} persists at all imbalances --- may be key to more accurately determining the boundary in parameter space between the regimes of dominant electron and ion heating, which could have important consequences for studies of the solar wind seeking to model its global properties. More broadly, the model for the electron-ion heating ratio proposed in \cref{sec:consequences_for_sw_heating} is necessarily incomplete since $Q_e/Q_i$ depends on the imbalance through the normalised cross-helicity, which, in the helicity-barrier regime, will itself be determined by the heating mechanisms that went into \cref{eq:heating_helicity_barrier}. Unfortunately, determining both $Q_e/Q_i$ and $\sigma_c$ self-consistently is a challenging task, given that capturing the dominant mechanisms of ion heating in the helicity-barrier regime requires resolving the full six-dimensional phase-space structure of the ion distribution function even before electron kinetics are added. Assuming that the effects of plasma echoes remain insignificant, the results of \cref{sec:landau_damping} suggest that the effects of electron kinetics across a range of betas and imbalances could be well-approximated by, e.g., a `Landau-fluid' closure \citep{hammett90,hammett92,hammett93,dorland93,beer96,snyder97,passot04,goswami05,passot17}. Using hybrid-kinetic simulations with such a closure for electrons \citep[see, e.g.,][]{finelli21} may thus be a promising path forward in the absence of full kinetic simulations. Despite these potential challenges, however, advancing our understanding of this `winner takes all' picture of heating that is emerging in the context of imbalanced Alfv\'enic turbulence is essential since it could have broad implications for plasma turbulence across a diverse array of astrophysical environments, including the solar wind, solar corona, accretion flows, and the intracluster medium.

%% Please use the acknowledgment and contribution environments. This will 
%% be anonomyized when the "anonymous" style option is used. 
\begin{acknowledgments}
The authors would like to thank B. D. G. Chandran, P. J. Ivanov, M. W. Kunz, A. A. Schekochihin, and M. Zhou for helpful discussions and suggestions at various stages of this project. The authors acknowledge the support of the Royal Society Te Ap\=arangi, through Marsden-Fund grant MFP-UOO2221 (TA and JS) and MFP-U0020 (RM). The work of RM was also supported in part by NASA grant 80NSSC24K0171. The work of TA was supported by the Laboratory Directed Research and Development (LDRD) Program at the Princeton Plasma Physics Laboratory for the U.S. Department of Energy under Contract No. DE-AC02-09CH11466. The United States Government retains a non-exclusive, paid-up, irrevocable, world-wide license to publish or reproduce the published form of this manuscript, or allow others to do so, for United States Government purposes. Computational support was provided by the New Zealand eScience Infrastructure (NeSI) high-performance computing facilities (funded jointly by NeSI's collaborator institutions and through the Ministry of Business, Innovation \& Employment's Research Infrastructure programme), as well as by the NASA High-End Computing (HEC) Program through the NASA Advanced Supercomputing (NAS) Division at Ames Research Center.
\end{acknowledgments}

\begin{contribution}
TA, RM, and JS jointly contributed to the research design and writing of the manuscript; RM developed the code and diagnostics; TA developed the theory, benchmarked the code, ran simulations, and performed the analysis of simulation outputs.

\end{contribution}

\appendix

\section{Convergence in number of evolved Hermite moments $\nhermite$}
\label{app:dependence_on_nhermite}
Within the numerical framework discussed in \cref{sec:numerical_setup}, the otherwise infinite hierarchy of Hermite moments represented by \cref{eq:gms_equation} is closed via truncation, i.e., setting $g_m = 0$ for $m > \nhermite$. Such a truncation is valid only if the amplitudes of modes with $m \lessapprox \nhermite$ are small enough that the absence of a mode at $m = M+1$ does not significantly affect the results; the hypercollisions \cref{eq:hypercollisions} are added in an attempt to ensure that this is the case. Here, using the simulations labelled `Hermite scan' in \cref{tab:simulation_parameters}, we investigate the convergence of various properties of the KREHM system of equations with the number of evolved Hermite moments $M$, finding that a relatively low number of evolved moments is required in order to reproduce the correct numerical behaviour. In order to cover the number of different regimes considered in this paper, we perform convergence tests at the smallest and largest values of $\beta_e$ used [$\beta_e (m_e/m_i)^{-1} = 1.0$ and $\beta_e (m_e/m_i)^{-1} = 550.9$, respectively] in both the balanced ($\sigma_\varepsilon = 0.0$) and imbalanced ($\sigma_\varepsilon = 0.8$) regimes. For each, $\nhermite$ is varied in steps of four over the range $4 \leqslant \nhermite \leqslant 32$. All of the following data presented has been averaged over the last 20\% of the simulation time.

\begin{figure}
	\centering
	\includegraphics[width=\figscale\textwidth]{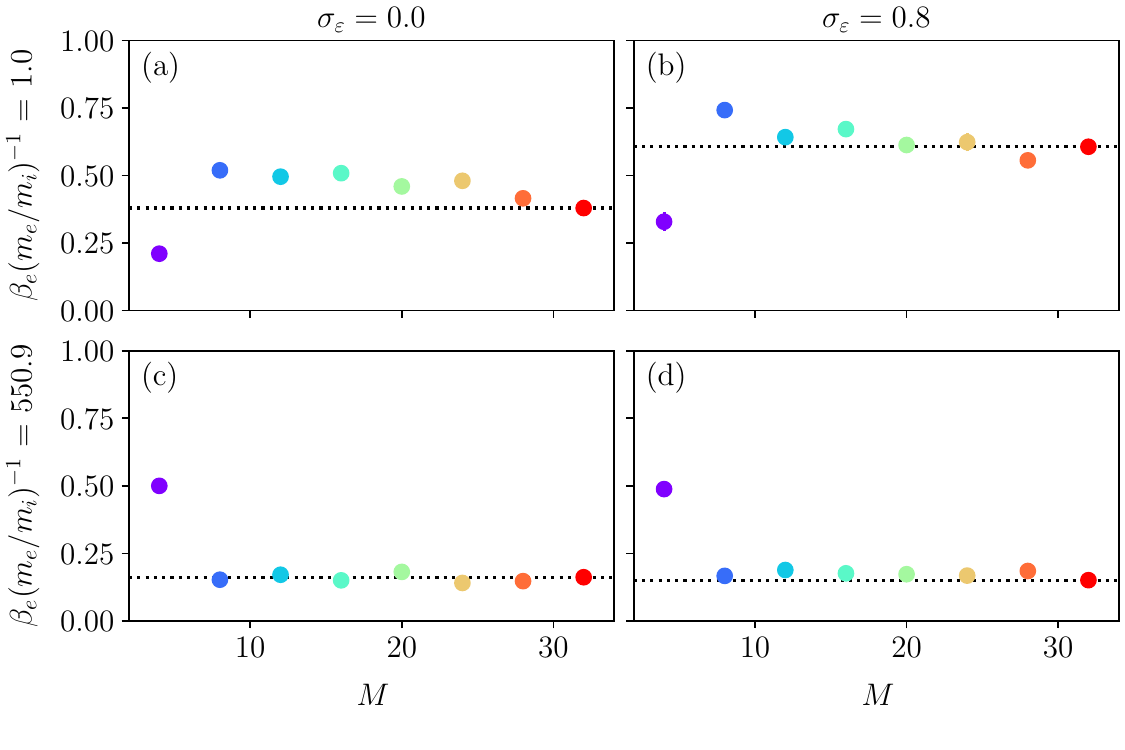}
	
	\caption[]{The parallel-dissipation rate $D_\parallel$ \cref{eq:parallel_diss}, normalised to the total energy flux $\fluxe$, as a function of $\nhermite$ for each set of parameters in the `Hermite scan' set of simulations from \cref{tab:simulation_parameters}. The values of the injection imbalance $\sigma_\varepsilon$ and normalised electron beta $\beta_e (m_e/m_i)^{-1}$ for each panel are indicated by the column and row headers, respectively. The horizontal dashed line indicates the value for $\nhermite = 32$.}
	\label{fig:hermite_scan_heating_rates}
\end{figure}

\begin{figure}
	\centering
	\includegraphics[width=\figscale\textwidth]{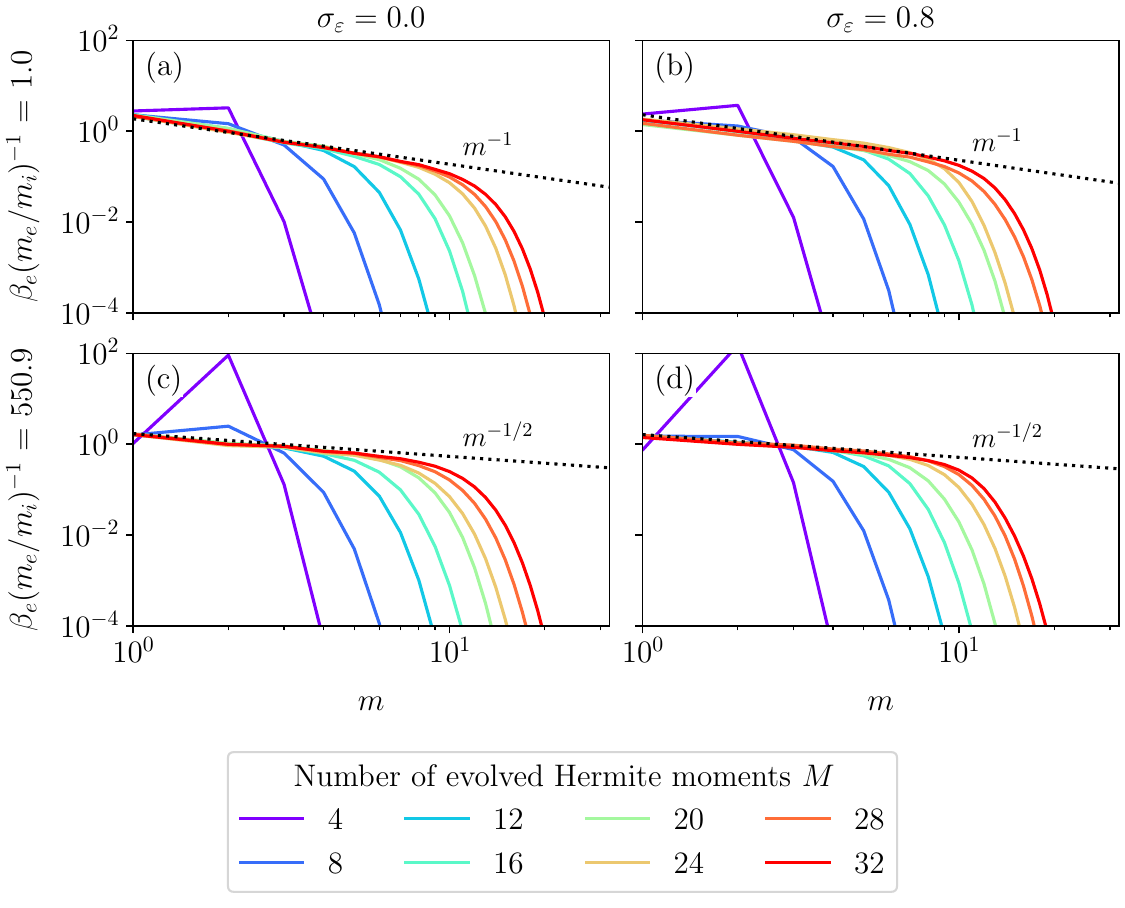}
	
	\caption[]{The one-dimensional Hermite spectra $E_m$ \cref{eq:hermite_spectrum} for the `Hermite scan' set of simulations from \cref{tab:simulation_parameters}, normalised to the value of the $\nhermite =32$ spectrum at $m=2$, with the colours indicating the value of $\nhermite$ for each spectrum. The values of the injection imbalance $\sigma_\varepsilon$ and normalised electron beta $\beta_e (m_e/m_i)^{-1}$ for each panel are indicated by the column and row headers, respectively. The dashed lines show approximate scalings of the spectra with $m$.}
	\label{fig:hermite_scan_hermite_spectra}
\end{figure}

As could have been anticipated, a value of $\nhermite=4$ is insufficient to allow the numerical scheme to behave properly, with those simulations significantly under- or over-estimating the parallel-dissipation rate \cref{eq:parallel_diss}, as can be seen from \cref{fig:hermite_scan_heating_rates}. This is a result of the improper behaviour of the one-dimensional Hermite spectrum \cref{eq:hermite_spectrum}, with the $m=2$ moment having a higher amplitude than the $m=1$ moment (see \cref{fig:hermite_scan_hermite_spectra}), likely due to the reflection of energy from the boundary in Hermite space provided by the truncation. As $\nhermite$ is increased, the Hermite spectra quickly start reproducing the same scaling as the $\nhermite = 32$ simulations before being cut off by the effects of the hypercollisions \cref{eq:hypercollisions}, with the parallel-dissipation rates behaving similarly. If the latter is taken as the primary metric for assessing convergence in $\nhermite$, then a value of $\nhermite = 12$ appears sufficient to accurately capture the same behaviour as at larger $\nhermite$. We note, however, that it is possible that the convergence observed here could change if the steepening of the Hermite flux $\Gamma_m$ observed at low beta [see \cref{fig:weakened_cascade_hermite}(c), red curve] meant that it became zero for some $m \leqslant \nhermite$, though this does not seem likely from the set of simulations conducted in this paper.

\section{Justification of $\kperpstar$ scaling}
\label{app:kperpstar}
In this appendix, we demonstrate that the scaling \cref{eq:kperpstar} corresponds to the point in perpendicular wavenumber space where co-propagating interactions begin to have an effect on the nonlinear dynamics. It is worth emphasising that the existence such interactions does immediately imply that a larger proportion of the injected energy flux will now be allowed to propagate to small scales. Rather, it signifies the breakdown of a specific property of RMHD: that the linear eigenmodes coincide with the functions that \textit{also} nullify the nonlinear interactions between co-propagating modes.

We take as our starting point the equations of FLR-MHD \citep{meyrand21}:
\begin{align}
	\frac{\rmd}{\rmd t} \taubar^{-1} \frac{e\phi}{T_{0e}} - \frac{c}{4 \pi e n_{0e}} \gpar \gperp^2 \Apar &= 0,  \label{eq:phi_equation_flrmhd} \\
	\frac{\rmd \Apar}{\rmd t} +  c \left(\frac{\partial \phi}{\partial z} + \gpar \taubar^{-1} \phi\right) & = 0 , \label{eq:apar_equation_flrmhd}
\end{align}
which represent the simplest model capable of capturing the dynamics of the helicity barrier. These can be obtained from \cref{eq:phi_equation} and \cref{eq:apar_equation} in the limit of $\kperp d_e \rightarrow 0$ while maintaining $\kperp \rhoi \sim 1$, corresponding to a regime where $\beta_e$ is sufficiently large that the effects of electron kinetics can be neglected ($m_e/m_i \ll \beta_e \ll 1$). Despite this only ever being an approximation, this is precisely the regime in which the scaling \cref{eq:kperpstar} was measured by \cite{meyrand21,squire22,squire23}.

Given that the break is located at scales $\kperp \rhoi \ll 1$ for sufficiently imbalanced (helicity-barrier-mediated) turbulence, it will be useful to work in terms of the standard Els\"asser potentials [cf. \cref{eq:thetapm}]:
\begin{align}
	\lim_{\kperp \rightarrow 0} \thetapm	= \zetapm \equiv \frac{c}{B_0} \phipot \mp \frac{\Apar}{\sqrt{4 \pi n_{0i}  m_i}},
	\label{eq:elsasser_potentials}
\end{align}
that are the stream functions for the RMHD Els\"asser fields $\vec{\zed}^\pm$ \citep{sch09}. Without loss of generality, \cref{eq:phi_equation_flrmhd} and \cref{eq:apar_equation_flrmhd} can then be recast in these variables as:
\begin{align}
	&\frac{\partial}{\partial t}\taubar^{-1} \zetapm \mp \frac{v_A}{2} \frac{\partial}{\partial z}\left[\left(1 + \taubar^{-1}\right)\taubar^{-1} \mp \rhos^2 \gperp^2 \right]\zetap+ \mp \frac{v_A}{2} \frac{\partial}{\partial z}\left[\left(1 + \taubar^{-1}\right)\taubar^{-1} \pm \rhos^2 \gperp^2 \right]\zetam  \nonumber \\
	&+ \frac{1}{4} \pbra{\zetap}{\left(\taubar^{-1} + \rhos^2 \gperp^2\right) \zetap} + \frac{1}{4}\pbra{\zetam}{\left(\taubar^{-1} + \rhos^2 \gperp^2\right) \zetam} + \frac{1}{4} \pbra{\zetap}{\left(\taubar^{-1} - \rhos^2 \gperp^2\right) \zetam} \nonumber \\
	&+ \frac{1}{4} \pbra{\zetam}{\left(\taubar^{-1} - \rhos^2 \gperp^2\right) \zetap} \mp \frac{\taubar^{-1}}{4} \pbra{\zetap - \zetam}{\left(1+\taubar^{-1}\right)\left(\zetap + \zetam\right)} = 0, \label{eq:flrmhd_elsasser}
\end{align}
where $\pbra{\dots}{\dots} = \ub_0 \cdot \left[\grad{(\dots)} \times \grad{(\dots)}\right]$ is the standard Poisson bracket. Though this representation is entirely equivalent to \cref{eq:phi_equation_flrmhd}-\cref{eq:apar_equation_flrmhd}, it has the advantage of explicitly separating the large-scale co-propagating ($\zetapm$ with $\zetapm$) and counter-propagating ($\zetapm$ with $\zetamp$) nonlinear interactions. 

Recalling \cref{eq:quasineutrality}, we can expand the $\taubar^{-1}$ operator in the long-wavelength limit as 
\begin{align}
	\taubar^{-1} = - \left(1+ \frac{3}{8} \rhoi^2 \gperp^2\right) \rhos^2 \gperp^2 + \dots,
	\label{eq:taubar_expansion}
\end{align}
from which it is straightforward to show that all co-propagating vanish to leading order, giving the equations of Els\"asser RMHD \citep{sch09,sch22}:
\begin{align}
	\frac{\partial }{\partial t}\gperp^2 \zetapm \mp v_A \frac{\partial }{\partial z} \gperp^2 \zetapm + \pbra{\zetamp}{\gperp^2 \zetapm} - \pbra{\partial_a \zetapm}{\partial_a \zetamp} = 0. 
	\label{eq:rmhd_elsasser}
\end{align}
At next order, one obtains modifications to both the linear terms and nonlinearities between counter-propagating fluctuations, but also additional nonlinear terms of the form
\begin{align}
	\pbra{\zetapm}{\rhoi^2 \gperp^4 \zetapm} \quad \text{or} \quad \gperp^2 \pbra{\zetapm}{\rhos^2 \gperp^2 \zetapm},
	\label{eq:extra_terms}
\end{align}
that describe co-propagating interactions. 
Formally, these terms are smaller than those in \cref{eq:rmhd_elsasser} by a factor of $\kperp^2 \rhoi^2 \ll 1$ if the turbulence is balanced, i.e. when $\zetap \sim \zetam$. At sufficiently high imbalance, however, it is possible for these terms to compete with those in \cref{eq:rmhd_elsasser}. Defining the characteristic Els\"asser amplitudes $\zetapms$ analogously to $\thetapms$ [see \cref{eq:thetapm_spectra_def}], we can, neglecting any possible anisotropy in the perpendicular plane, estimate the size of the nonlinearities in \cref{eq:rmhd_elsasser} and \cref{eq:extra_terms} as, respectively,
\begin{align}
	\pbra{\zetamp}{\gperp^2 \zetapm} \sim \kperp^4 \zetaps \zetams, \quad   \pbra{\zetapm}{\rhoi^2 \gperp^4 \zetapm}  \sim \kperp^4 (\kperp \rhoi)^2 \left(\zetapms\right)^2.
	\label{eq:nonlinear_estimate}  
\end{align}
Noting that the nonlinearities involving co-propagating $\zetam$ fluctuations will always be subdominant for $\zetam \ll \zetap$, we can estimate, by balancing the two terms in \cref{eq:nonlinear_estimate}, the scale at which the co-propagating $\zetap$ fluctuations nonlinearly compete with the counter-propagating ones:
\begin{align}
	\kperp \rhoi \sim \left(\frac{\zetams}{\zetaps}\right)^{1/2} \sim \left(\frac{W^-}{W^+}\right)^{1/4}.
	\label{eq:kperpstar_appendix}
\end{align}
The final estimate in \cref{eq:kperpstar_appendix} assumes that both fields have approximately the same slope on scales above \cref{eq:kperpstar_appendix}, which seems to be the case in simulations of strongly-imbalanced turbulence \citep[see, e.g.,][]{beresnyak09slopes,meyrand21,sch22}. Finally, noting that $\sigma_c \approx 1 - W_-/W_+$ at large imbalance, this estimate becomes \cref{eq:kperpstar}. 

It would thus appear that the scaling of the break position with imbalance is correlated with the importance of nonlinear interactions between co-propagating fluctuations of the larger field within the turbulence. If true, this has interesting consequences for studies of imbalanced RMHD turbulence. The validity of the RMHD system of equations \cref{eq:rmhd_elsasser} is predicated on the assumption that there is no special scale at $\kperp \rhoi \ll 1$, since the equations are manifestly scale invariant. However, the existence of some special scale, particularly one which shifts to longer wavelengths as the imbalance increases, restricts the range of perpendicular wavenumbers where RMHD remains valid. While weak scaling of \cref{eq:kperpstar_appendix} with imbalance suggests this restriction of the range of validity of RMHD will be relatively minor, the existence of this scale has important consequences for the thermodynamics of heating in imbalanced turbulence due to the role that it plays in the evolution of the helicity-barrier-dominated state that was considered in \cref{sec:landau_damping}, and the fact that it could lead to an improved phenomenology of the transition range.

%% For this sample we use BibTeX plus aasjournals.bst to generate the
%% the bibliography. The sample7.bib file was populated from ADS. To
%% get the citations to show in the compiled file do the following:
%%
%% pdflatex sample7.tex
%% bibtext sample7
%% pdflatex sample7.tex
%% pdflatex sample7.tex

\bibliography{bibliography.bib}{}
\bibliographystyle{aasjournal_v6.3.1}

%% This command is needed to show the entire author+affiliation list when
%% the collaboration and author truncation commands are used.  It has to
%% go at the end of the manuscript.
%\allauthors

%% Include this line if you are using the \added, \replaced, \deleted
%% commands to see a summary list of all changes at the end of the article.
%\listofchanges

\end{document}